\documentclass[]{acmart}
\AtBeginDocument{%
  }
\usepackage{algorithm}
\usepackage{enumitem}
\usepackage{tabularx}
\usepackage{xcolor}
\usepackage{subcaption}

\setcopyright{cc}
\setcctype{by}
\acmJournal{PACMHCI}
\acmYear{2026} \acmVolume{10} \acmNumber{4} \acmArticle{EICS027}
\acmMonth{6} \acmDOI{10.1145/3816779}
\acmConference[EICS]{The 18th ACM SIGCHI Symposium on Engineering Interactive Computing Systems}{30 June - 3 July 2026}{Patras, Greece}

\begin{document}

\title{Continuous Behavioral Synthesis for Adaptive Health Dashboards: An LLM-Mediated Architecture Integrating Explicit Preference, Spatial Reorganization, and Attention Allocation Signals}

\author{Tiziano Santilli}
\email{tisa@mmmi.sdu.dk}
\affiliation{%
  \institution{University of Southern Denmark, MMMI Software Engineering}
  \city{Odense}
  \country{Denmark}}

\author{Mina Alipour}
\email{mial@mmmi.sdu.dk}
\affiliation{%
  \institution{University of Southern Denmark, MMMI Software Engineering}
  \city{Odense}
  \country{Denmark}}

\author{Mahyar T. Moghaddam}
\email{mtmo@mmmi.sdu.dk}
\affiliation{%
  \institution{University of Southern Denmark, MMMI Software Engineering}
  \city{Odense}
  \country{Denmark}}

\renewcommand{\shortauthors}{Santilli et al.}

\begin{abstract}
The engineering of adaptive user interfaces has traditionally relied on either rule-based systems encoding designer intuitions about user needs or machine learning approaches requiring substantial historical data before achieving effective personalization. We present a technical architecture that leverages Large Language Models as behavioral synthesis engines to enable immediate adaptation from sparse, heterogeneous user signals. Our system integrates three distinct behavioral channels, i) explicit micro-feedback on individual interface elements, ii) spatial priority inferred from manual widget reorganization through drag-and-drop interaction, iii) and attentional investment measured through dwell time during hover events, within a structured prompt engineering framework that continuously regenerates dashboard layouts while maintaining explanatory coherence. The architecture addresses the technical challenge of translating low-level interaction patterns into high-level design decisions through a layered prompt construction methodology that separates temporal context determination, behavioral signal extraction, explicit preference enforcement, and user profile synthesis. The approach combines manually specified behavioral interpretations and temporal heuristics with LLM-mediated synthesis, enabling the reconciliation of multiple simultaneous signals that would be difficult to encode through explicit rules alone.

We demonstrate the system through an instantiation in the personal health monitoring domain, including an analytical evaluation of adaptation behavior across multiple scenarios and a working implementation managing fourteen distinct health metrics across seven widget visualization modalities.
The evaluation compares profile-driven initialization, multi-signal behavioral adaptation, and presents the resulting interfaces through representative post-adaptation screenshots.
The analytical evaluation shows that the system preserves explicit user constraints, keeps user-prioritized metrics in prominent positions, expands widgets that receive sustained attention.
The technical contribution comprises the multi-modal behavioral aggregation strategy, the structured LLM prompt engineering approach for maintaining design consistency across regeneration cycles, and the explainability generation mechanism that exposes adaptation rationale to end users. Our work provides a reproducible engineering approach for building LLM-powered adaptive interfaces that can be generalized beyond health dashboards to any domain requiring continuous interface personalization from heterogeneous user behavior.
A current limitation is that, while the system can infer and act on behavioral signals, it does not yet incorporate mechanisms to independently verify whether adaptations improve user experience without additional feedback signals.
\end{abstract}



\maketitle

\section{Introduction}

The fundamental challenge in engineering adaptive user interfaces lies in the translation problem: converting low-level user interactions (clicks, drags, hovers, explicit ratings) into high-level interface design decisions about information architecture, visual hierarchy, and presentation modality. Traditional approaches have addressed this challenge through two primary strategies~\cite{Rothrock2002,Lavie2010}, each with significant engineering limitations. Rule-based adaptive systems~\cite{Gajos2008,Bunt2007} encode designer intuitions about how specific user characteristics or behaviors should map to interface modifications, but these rigid mappings struggle to handle the combinatorial complexity of multiple simultaneous user signals and often produce brittle systems that fail gracefully under unexpected interaction patterns. Machine learning approaches~\cite{Findlater2009}, conversely, can discover nuanced relationships between user behavior and interface effectiveness, but they require substantial training data before achieving useful personalization and typically operate as black boxes that provide no rationale for their adaptation decisions.

The recent emergence of Large Language Models as general-purpose reasoning engines presents a novel engineering opportunity: using LLMs not as endpoint generators but as behavioral synthesis components within adaptive interface architectures. Unlike traditional adaptive systems that must be explicitly programmed with every behavioral mapping or trained on extensive datasets, LLMs possess embedded knowledge about human behavior, interface design principles, and domain-specific conventions that can be activated through structured prompting. However, realizing this potential requires careful engineering to address several technical challenges. First, LLMs must receive behavioral signals in a structured format that preserves semantic meaning while remaining interpretable by the language model. Second, the prompt construction mechanism must maintain consistency across multiple regeneration cycles to prevent erratic interface changes that would confuse users. Third, the system must generate not only interface specifications but also human-readable explanations of why specific adaptations were chosen, supporting user trust and enabling corrective feedback.

The architecture is intentionally hybrid. Several adaptation primitives are manually specified by the designers: the coarse temporal phases, the five-hundred-millisecond dwell threshold, the interpretation of explicit likes and dislikes, and the convention that user-initiated repositioning toward top positions expresses higher informational priority. The LLM should therefore not be understood as replacing all rules. It performs contextual synthesis across multiple simultaneously active signals that would otherwise require a large and brittle rule table. In practice, the LLM resolves conflicts among profile goals, temporal context, explicit preferences, spatial priorities, and detailed-attention cues; chooses a coherent combination of widgets, sizes, and chart modalities; and generates a user-facing explanation of why that configuration was produced. The system is therefore best described as a hybrid synthesis architecture rather than a purely learned adaptive system. This aligns with mixed-initiative interface research, which treats effective adaptation as a negotiated balance between system initiative, user control, and interpretable rationale rather than as full automation alone \cite{Bunt2007}.

We present a technical architecture that addresses these challenges through a multi-layered approach to behavioral synthesis. The core innovation lies in treating the LLM as a continuous synthesis engine that periodically integrates accumulated behavioral signals rather than as a one-shot interface generator invoked once during system initialization. Our architecture maintains three distinct behavioral tracking channels that operate asynchronously with respect to each other and to the adaptation cycle itself. Explicit micro-feedback, implemented through lightweight like and dislike buttons attached to each dashboard widget, provides unambiguous preference signals about specific interface elements. Spatial reorganization tracking, implemented through HTML5 Drag and Drop API event handlers, captures user-initiated changes to widget positioning and interprets these actions as implicit priority signals about information importance. Attention allocation monitoring, implemented through hover event duration tracking with noise filtering for incidental mouse movements, reveals which interface elements users invest cognitive resources in understanding, suggesting either confusion requiring simplification or genuine interest warranting expanded detail.

The technical challenge of synthesizing these heterogeneous signals into coherent interface adaptations is addressed through a structured prompt engineering methodology. Rather than simply concatenating behavioral data and asking the LLM to "improve the interface," our approach constructs prompts with four distinct conceptual sections, each serving a specific role in constraining the LLM's generation process. Temporal context framing provides time-of-day awareness, enabling the system to prioritize different health metrics during morning preparation periods, active daytime hours, and evening reflection phases. Behavioral signal presentation formats the three tracking channels into natural language descriptions that explicitly state the semantic interpretation of observed patterns, such as "The user has dragged steps and sleep quality to the top three positions, indicating these metrics represent current priorities." Explicit preference enforcement translates accumulated like and dislike signals into hard constraints that the LLM must satisfy, preventing the generation of layouts containing disliked metrics regardless of other considerations. User profile synthesis incorporates static demographic information, health goals, and interface preferences as additional context for personalization decisions.

This architectural approach yields several technical advantages over alternative adaptive interface designs. By operating the LLM in a synthesis mode rather than a generation mode, the system can begin adapting immediately from the first user interaction rather than requiring an initialization period to gather sufficient training data. By maintaining separation between behavioral tracking and adaptation logic, the architecture supports extensibility-new behavioral channels can be added by implementing additional tracking modules and extending the prompt construction logic without modifying the core adaptation engine. By generating structured JSON output with explicit widget specifications rather than generating interface code directly, the system preserves separation of concerns between adaptive logic and rendering implementation. By producing explanatory text alongside interface specifications, the architecture enables transparency about adaptation decisions without requiring separate explanation generation machinery.

We demonstrate this architectural approach through a complete prototype in the personal-health dashboard domain. The prototype is not presented as a validated claim that the generated adaptations improve usability, adherence, or health outcomes. Instead, it serves as a concrete instantiation of three engineering design proposals: multi-channel behavioral aggregation, structured LLM-mediated synthesis, and integrated explainability generation. The health-dashboard setting is an informative testbed because personal informatics and wellbeing applications routinely combine heterogeneous measurements, repeated short-term interpretation, and individualized goal structures, while also requiring feedback to fit everyday routines and shifting contexts. Prior work on personal informatics and self-tracking has similarly shown that such systems are most useful when they support reflection on varied data types without assuming that all users need the same metrics, the same visual structure, or the same cadence of interpretation \cite{chopra2025engagements}.

The remainder of this technical note proceeds as follows. Section 2 reviews related technical work in adaptive user interfaces, LLM-based interface generation, and health information presentation. Section 3 presents the detailed system architecture, including the multi-modal behavioral tracking subsystem, the prompt engineering methodology, and the rendering pipeline. Section 4 describes the working implementation, including technical details necessary for reproduction. Section 5 demonstrates system capabilities through concrete usage scenarios. Section 6 discusses technical limitations, generalization potential, and future engineering challenges. Section 7 concludes.

The supplemental materials, comprising the source code, user data, user dashboard screenshot, prompt list, and installation guide (with accompanying video), are available at this link\footnote{https://anonymous.4open.science/r/health-dynamic-dashboard-4317}.

\section{Related Technical Work}

\subsection{Adaptive User Interface Engineering}

The engineering of adaptive user interfaces represents a long-standing research challenge within interactive systems. Early model-based approaches~\cite{Gajos2008,Rothrock2002} attempted to formalize the adaptation process through explicit user models that could be queried to determine appropriate interface configurations. These systems typically employed rule-based logic to map user characteristics to interface parameters, enabling reasoning about adaptations but requiring extensive manual specification of adaptation rules. The brittleness of hand-crafted rules led to increased interest in machine learning approaches that could discover adaptation strategies from data. Reinforcement learning frameworks~\cite{Teso2024} model interface adaptation as a sequential decision problem where the system learns optimal action policies by observing user response to interface changes over time. While theoretically appealing, these approaches face the cold start problem-they require extensive interaction history before producing effective adaptations-and the credit assignment problem-determining which specific interface changes contributed to observed improvements in user behavior.

More recent work has explored hybrid approaches that combine learning with designer-specified structure. Context-aware adaptive systems~\cite{Bunt2007} use sensors to detect user environment and activity, adapting interfaces based on inferred situations rather than learned policies. These systems demonstrate that adaptation can be effective even without user-specific learning if sufficient contextual information is available. However, they typically focus on relatively coarse adaptations like switching between interface modes rather than fine-grained personalization of information presentation.

\subsection{Large Language Models in Interface Generation}

The application of Large Language Models to user interface generation has emerged as a distinct research area within HCI. Initial work~\cite{Tian2023,Chen2024} focused on generating interface implementation code from natural language descriptions, treating LLMs as code synthesis tools that translate high-level intent into low-level implementation. These systems demonstrate impressive capabilities in producing functional interfaces but operate in a one-shot generation mode unsuited to continuous adaptation. More recent research~\cite{Wu2023,Jiang2024} has explored using LLMs as design reasoning engines that can evaluate interface alternatives and explain design decisions. This work suggests that LLMs embed substantial knowledge about interface design principles and can articulate rationale for design choices in human-readable form.

Particularly relevant to our work is recent exploration~\cite{Wang2025,Li2025} of LLMs for accessibility adaptation, where language models dynamically adjust interface complexity and modality based on user accessibility needs. These systems demonstrate that LLMs can reason about appropriate adaptations given user context, but they typically rely on explicit user profiles rather than learned behavioral patterns. Our contribution extends this line of work~\cite{Kim2024} by developing an architecture for continuous behavioral learning through LLM-mediated synthesis.

\subsection{Health Dashboard Design}

Personal health dashboards present unique technical challenges~\cite{Bentley2013,Choe2014} in information architecture and visualization selection. Users must interpret diverse data types~\cite{Rooksby2014,Epstein2015} ranging from single measurements to temporal trends to compositional breakdowns, each requiring different visualization approaches. Existing health platforms typically employ static dashboard designs that present all available metrics uniformly, leading to information overload and reduced engagement over time. Research on personalized health dashboards has explored rule-based customization~\cite{Mamykina2016,West2013} where users manually configure which metrics to display, but studies~\cite{Lazar2015,Fritz2014} indicate that users struggle to anticipate their own future information needs and that optimal dashboard configurations vary based on temporal context and current health goals.

Recent work~\cite{Consolvo2009,Fogg2009} has begun exploring adaptive health visualizations that adjust automatically based on user behavior. These systems typically employ simpler adaptation strategies like promoting frequently viewed metrics or highlighting abnormal measurements, but they lack sophisticated reasoning about why specific metrics might be relevant in particular contexts. Our system contributes to this area by demonstrating how behavioral synthesis through LLMs can enable more nuanced health dashboard adaptation.

\section{System Architecture}

\subsection{Architectural Overview}

\begin{figure*}[h]
\centering
\includegraphics[width=0.5\textwidth]{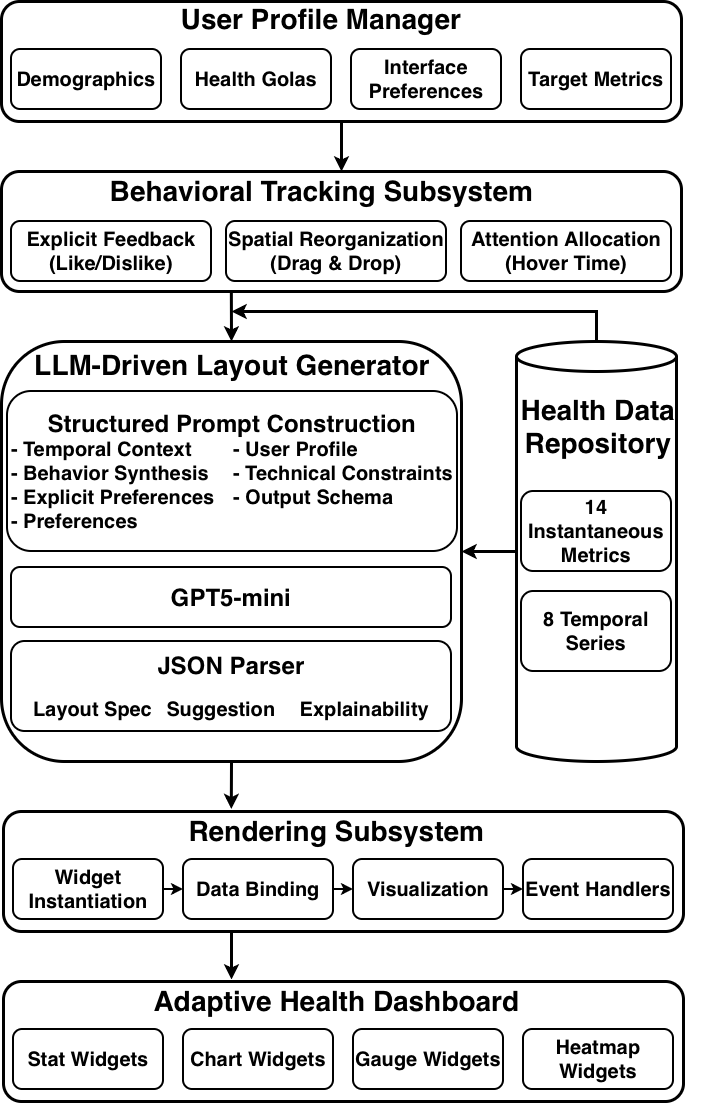}
\caption{System architecture showing the four primary subsystems and their interaction patterns. The Behavioral Tracking Subsystem monitors user interactions across three independent channels (explicit feedback, spatial reorganization, attention allocation). The LLM-Driven Layout Generator synthesizes accumulated behavioral signals through structured prompt construction, invokes GPT-5-mini to generate dashboard specifications, and parses the JSON response. The Rendering Subsystem transforms abstract widget specifications into concrete interface implementations. The Adaptive Dashboard displays the generated layout while capturing new user interactions that feed back into the behavioral tracking system, completing the continuous adaptation cycle.}
\label{fig:architecture}
\Description{System Architecture}
\end{figure*}

The system architecture (Figure~\ref{fig:architecture}) comprises four primary subsystems
 that operate in a coordinated cycle to produce continuously adaptive health dashboards. The Behavioral Tracking Subsystem monitors user interactions across multiple channels, maintaining timestamped logs of explicit feedback events, spatial reorganization actions, and attention allocation patterns. These logs accumulate asynchronously with respect to the adaptation cycle, enabling the system to learn from user behavior continuously rather than only during discrete adaptation periods. The Profile Management Subsystem maintains user demographic information, health goals, and interface preferences, providing stable context that complements dynamic behavioral signals. The Layout Generation Subsystem receives behavioral logs and profile data, constructs structured prompts that synthesize this information into actionable design instructions, invokes the Large Language Model to generate dashboard specifications, and parses the resulting structured output into internal interface representations. The Rendering Subsystem transforms these abstract specifications into concrete HTML and JavaScript implementations, handling widget instantiation, data binding, event handler attachment, and visual layout in a CSS grid structure. These four subsystems communicate through well-defined interfaces using JSON as the primary data interchange format, enabling modular implementation and facilitating system extension.

\subsection{Multi-Modal Behavioral Tracking}

The behavioral tracking subsystem implements three distinct monitoring channels that capture qualitatively different aspects of user engagement with the dashboard interface. Each channel operates through instrumented event handlers attached to the dashboard DOM structure, recording interaction patterns and maintaining cumulative statistics that inform subsequent adaptations.

The explicit feedback channel implements a micro-survey approach embedded directly within each dashboard widget. Two lightweight button elements rendered in each widget's header region provide binary valence feedback: a positive affordance for expressing satisfaction with the widget's presence and a negative affordance for indicating the widget should be excluded from future layouts. These affordances employ minimalist iconography to avoid visual clutter while remaining semantically clear through hover tooltips. When a user activates either affordance, the system records a timestamped feedback event specifying the target metric identifier, the expressed sentiment, and the current dashboard configuration context. These events accumulate in the user's persistent history log, enabling the system to track preference evolution over time and to detect preference consistency or volatility. Crucially, the architecture treats these explicit signals as hard constraints during layout generation: disliked metrics are strictly excluded from generated layouts regardless of other considerations, while liked metrics receive prioritization for prominent placement. This constraint enforcement prevents the frustrating user experience pattern where an adaptive system repeatedly surfaces content that users have explicitly rejected.

The spatial reorganization channel leverages the HTML5 Drag and Drop API to detect and interpret user-initiated changes to widget positioning within the dashboard grid. The implementation attaches dragstart, dragend, dragover, and drop event handlers to the dashboard container element and all widget child elements. When a user initiates a drag operation by clicking and holding a widget, the dragstart handler marks the target element as currently dragging and enables visual feedback through CSS class modification. The dragover handler executes continuously during the drag operation, computing the geometric relationship between the cursor position and all non-dragging widgets to determine appropriate insertion points. This computation handles the two-dimensional grid layout by projecting cursor coordinates onto widget bounding boxes and identifying the nearest valid drop position according to CSS grid semantics. When the user releases the mouse button, the dragend handler commits the reorganization by updating the underlying data structure ordering to match the new visual arrangement, persisting this change to the database, and updating local state to ensure consistency.

The technical contribution of this channel lies not merely in enabling drag-and-drop reorganization (a standard interface capability) but in the structured semantic interpretation and integration of these actions within a multi-signal adaptation pipeline. Interpreting top-positioned items as more important is a well-established interface convention; however, in this system, such spatial manipulations are explicitly captured and operationalized as machine-readable signals that contribute to downstream adaptation decisions.
When a user drags a widget to a new position, particularly to a prominent position near the top-left origin of the grid, the system interprets this action as an implicit priority signal indicating heightened importance of the represented metric. This interpretation persists and is later incorporated into prompt construction, where it is communicated explicitly to the LLM: “The user has manually positioned these metrics at the top of their dashboard, indicating current high priority. While you may adjust their visual presentation or size, maintain their prominence in the layout to respect this expressed preference.”
Crucially, drag-and-drop is not treated as a complete adaptation mechanism on its own. Instead, it is considered one interpretable signal among several, alongside explicit feedback, temporal context, and attention-based dwell signals. The role of the LLM in this context is to reconcile these potentially competing signals into a coherent layout specification. For example, a metric positioned prominently by the user may still be expanded, re-visualized, or contextually grouped differently if sustained attention or temporal relevance suggests a need for richer representation. This semantic bridging from low-level spatial manipulation to high-level design reasoning enables the system to preserve user intent while still supporting adaptive reconfiguration beyond what fixed-rule mappings would allow.

The attention allocation channel monitors hover events to infer which dashboard elements users invest cognitive effort in understanding. The implementation attaches mouseenter and mouseleave event handlers to each widget element, recording entrance timestamps and computing dwell duration upon exit. A critical technical detail involves filtering transient hovers resulting from incidental mouse movement while navigating to other interface regions. The system implements a temporal threshold mechanism that discards dwell periods shorter than five hundred milliseconds, retaining only sustained attention events likely to reflect genuine cognitive engagement. Qualified dwell events accumulate in a session-level interest map that maintains cumulative attention duration for each metric across the current session. This map provides a complementary signal to spatial reorganization: while drag-and-drop reveals deliberate prioritization decisions, dwell time reveals investigative attention that may indicate either confusion requiring simplification or genuine interest warranting expanded detail.

The prompt engineering logic interprets prolonged dwell time as a request for increased detail rather than an indicator of confusion, based on the design principle that users will disengage from confusing content rather than continuing to study it. Consequently, metrics accumulating substantial dwell time receive adaptation instructions to expand their presentation: "The user has spent significant time examining this metric, suggesting deep interest in understanding its patterns. Expand this widget to wide size or convert from numerical summary to detailed chart visualization to support this analytical engagement." This interpretation could be refined through additional behavioral signals-for example, differentiating confused hovering characterized by cursor movement within the widget from analytical hovering characterized by stable cursor positioning-but the current implementation employs the simpler sustained-attention interpretation with successful results.

The interpretation of dwell time should be understood as a heuristic approximation rather than a definitive inference of user intent. Cursor and hover behavior can serve as lightweight proxies for attention, and prior work suggests that such signals provide useful indications of what users are examining \cite{huang2011no}. At the same time, these signals remain inherently indirect: prolonged hover may reflect analytical interest, hesitation, confusion, accessibility-related reading patterns, or transient disengagement.
Accordingly, the current system treats dwell time as a soft signal that suggests a potential need for increased detail or alternative representation, rather than as conclusive evidence of user preference or experience quality. This design choice allows the architecture to remain responsive to emerging patterns of engagement while avoiding overcommitment to any single behavioral interpretation. It also leaves room for future extensions in which additional signals or feedback mechanisms can refine these inferences and improve adaptation reliability.

\subsection{Structured Prompt Engineering for Behavioral Synthesis}

The prompt construction mechanism is the core architectural device through which the system translates heterogeneous observations into a constrained design request. The prompting strategy is presented here as an engineering design proposal embodied in the prototype rather than as a separately validated scientific claim. Prompting choices strongly affect controllability, transparency, and output consistency in LLM-based systems, and recent HCI work has shown that structured prompt decomposition and prompt-based prototyping are useful because they make model behavior easier to debug, compare, and reason about \cite{Wu2023, Jiang2024}. The prompt used in this system follows that logic by externalizing the adaptation rationale into explicit sections rather than relying on the model to infer behavioral semantics from raw logs alone.

The temporal section of the prompt is grounded in a lightweight, theory-informed assumption that the relevance of health information changes across the day. The three temporal phases are not presented as a complete behavioral theory. Rather, they function as interpretable domain heuristics informed by research showing that time of day can affect cognitive performance, that behavior-change technologies must fit long-term everyday use, and that context-sensitive repeated assessment is often necessary to interpret behavior ecologically rather than retrospectively \cite{folkard1980circadian, Consolvo2009, shiffman2008ecological}. In practical terms, morning layouts can foreground readiness and recovery, mid-day layouts can foreground activity and intervention opportunities, and evening layouts can foreground closure, reflection, and next-day preparation.

To improve readability and reproducibility, prompt excerpts are visually separated from the running text and presented in shaded boxes. Each box contains only the subsection currently under discussion, for example, the temporal framing fragment when phase-aware prompting is introduced, and the explicit-feedback fragment when preference constraints are described. This selective presentation keeps the narrative focused while making individual components of the prompt structure explicit and inspectable.

\begin{table*}[t]
\centering
\caption{Hierarchical Prompt Structure for LLM-Mediated Behavioral Synthesis}
\label{tab:prompt-structure}
\begin{tabular}{@{}p{3cm}p{4cm}p{7cm}@{}}
\toprule
\textbf{Prompt Section} & \textbf{Information Source} & \textbf{Purpose and Content} \\
\midrule
Temporal Context & System clock mapped to predefined phase rules & Establishes time-of-day awareness. Specifies current phase (morning/day/evening) and associated design strategy emphasizing appropriate metric types. Grounds adaptation in behavioral psychology principles about information salience timing. \\
\addlinespace
Behavioral Context & Three tracking channel logs aggregated across session history & Synthesizes observed interaction patterns into semantic interpretations. Translates drag-drop positioning into priority signals, dwell time into interest indicators, and interaction frequency into engagement metrics. Bridges low-level events to high-level design implications. \\
\addlinespace
Explicit Feedback & Accumulated like/dislike preference events & Enforces hard constraints on layout generation. Formats approved metrics as must-include requirements and rejected metrics as must-exclude prohibitions. Ensures user agency by making explicit preferences non-negotiable regardless of other considerations. \\
\addlinespace
User Profile & Questionnaire responses and demographic data & Provides stable personalization context beyond immediate behavior. Encodes health goals, current metrics and targets, activity level, stress level, dietary focus, and interface preferences (accessibility, density, aesthetic, visualization style). Enables goal-aware adaptation. \\
\addlinespace
Technical Constraints & System capabilities enumeration & Prevents LLM hallucination of non-existent data sources or invalid widget-data combinations. Explicitly lists available metrics by category and specifies compatibility rules (e.g., chart widgets require temporal data sources, stat widgets require instantaneous measurements). \\
\addlinespace
Output Schema & JSON structure specification & Defines required output format with two components: suggestion object containing three explanation tiers (actionable text, scientific nudge, meta-reasoning) and layout array containing widget specifications with type, title, size, dataSource, and chartType fields. Enables deterministic parsing. \\
\bottomrule
\end{tabular}
\end{table*}

\begin{table}[t]
\centering
\caption{Temporal Adaptation Strategies Encoded in Prompt Engineering}
\label{tab:temporal-strategies}
\begin{tabular}{@{}p{2.5cm}p{2cm}p{5.5cm}@{}}
\toprule
\textbf{Temporal Phase} & \textbf{Time Window} & \textbf{Adaptation Strategy} \\
\midrule
Morning Preparation & 5:00--11:00 & Prioritize readiness indicators: sleep quality assessment, heart rate variability as recovery metric, daily goal preview. Suppress premature metrics like calories burned or exercise completion that lack meaning early in day. \\
\addlinespace
Active Day & 11:00--18:00 & Emphasize real-time activity monitoring: step count with progress toward goal, stress level tracking enabling intervention, hydration reminders. Promote actionable feedback supporting behavioral correction during active hours. \\
\addlinespace
Evening Reflection & 18:00--5:00 & Focus on closure and achievement: daily totals for steps and calories, recovery score projection, wind-down recommendations. Suppress forward-looking metrics replaced by retrospective summaries and next-day preparation suggestions. \\
\bottomrule
\end{tabular}
\end{table}

The morning phase prioritizes readiness indicators including sleep quality, heart rate variability as a recovery metric, and progress toward daily goals, while explicitly suppressing metrics like calories burned that lack meaning early in the day. The active phase emphasizes real-time activity monitoring including step counts, stress level indicators, and hydration reminders that support behavioral correction during the day. The evening phase focuses on closure and reflection, highlighting daily achievement summaries, recovery scores, and wind-down recommendations that support healthy evening routines. The prompt communicates these strategies explicitly: "Time of Day: MORNING (Preparation Phase). Strategy: Focus on READINESS. Prioritize Sleep Quality, HRV, and Today's Goal. Hide 'Calories Burned' (too early)." This temporal framing provides the LLM with domain-appropriate design guidance while maintaining flexibility about specific implementation choices.

The behavioral context section synthesizes the three tracking channels into natural language descriptions that make the semantic meaning of observed patterns explicit to the LLM. For spatial reorganization, the system extracts the top three metrics from the current dashboard layout configuration, formats their identifiers into a comma-separated list, and constructs a behavioral interpretation statement: "OBSERVED BEHAVIOR: The user previously dragged these metrics to the TOP of their dashboard: [vo2\_max, sleep\_quality, steps]. IMPLICIT PRIORITY: These are currently the most important metrics to the user. ADAPTATION INSTRUCTION: You should keep these high-priority metrics near the top. You may change their visual style (for example from 'stat' to 'chart') to give more detail, but do not bury them." This structure explicitly bridges the gap between interaction pattern and design implication, enabling the LLM to understand not just what happened but why it matters for interface design.

For attention allocation, the system sorts the session interest map by cumulative dwell duration, extracts the top three metrics, formats them with their associated attention duration in seconds, and constructs an interpretation grounded in cognitive psychology: "OBSERVED MICRO-INTERACTIONS (Dwell Time): The user spent the most time hovering over: [vo2\_max (twelve seconds), heart\_rate\_variability (eight seconds), asymmetric\_walk (five seconds)]. PSYCHOLOGICAL INTERPRETATION: The user is deeply investigating these metrics. ADAPTATION INSTRUCTION: These widgets should be expanded to 'wide' size or converted to detailed 'charts' to satisfy the user's need for detail. Do not simplify them." Again, the prompt makes the reasoning explicit rather than expecting the LLM to infer appropriate design responses from raw behavioral data.

For explicit feedback, the system partitions the accumulated preference history into like and dislike sets, formats each set as a comma-separated metric list, and formulates these as hard constraints: "EXPLICIT FEEDBACK (User Voice): LIKED Metrics (Must Include): [steps, sleep\_quality, vo2\_max]. DISLIKED Metrics (Must Remove): [calories\_burned, body\_fat\_percentage]." The use of imperative language ("Must Include," "Must Remove") signals to the LLM that these represent non-negotiable requirements rather than soft preferences to be balanced against other considerations.

The user profile section incorporates stable demographic and preference information to personalize adaptations beyond immediate behavioral signals. This section encodes health goals (weight loss, fitness improvement, stress management, general wellness), current metrics including weight and target weight, activity level self-assessment, dietary focus, stress level, and interface preferences including visual accessibility needs, desired information density, aesthetic preferences, and preferred data visualization style. The prompt formats this information in a structured list: "USER PROFILE: Goal: weight loss. Metrics: Weight eighty-five kg (Target: seventy-five kg). Activity: moderate (one to three workouts per week). Stress Level: high. Complexity: cockpit mode (show me everything). UI Vibe: professional. Preferred Data Style: visual charts." This contextual information enables the LLM to make design decisions aligned with user goals and preferences even in the absence of specific behavioral signals. For example, users with weight loss goals reliably receive weight tracking widgets regardless of whether they have explicitly interacted with these widgets, based on the goal-metric relevance reasoning the LLM can perform.

Following these four contextual sections, the prompt specifies technical constraints necessary to prevent LLM hallucination of non-existent data sources or widget types. The system enumerates all available data sources by category, listing the fourteen instantaneous health metrics the system can display as stat or gauge widgets and the eight temporal data series available for chart widgets. The prompt explicitly warns against mismatches: "CRITICAL DATA SOURCE RULES: CHART WIDGETS require dataSource from: [steps\_weekly, sleep\_weekly, heart\_rate\_weekly, ...]. STAT or GAUGE WIDGETS require dataSource from: [today\_steps, today\_resting\_hr, today\_hr\_variability, ...]. NEVER use type 'chart' with a user\_stats dataSource (for example, today\_resting\_hr). If showing today\_steps, today\_resting\_hr, etc., use type 'stat', NOT type 'chart'." These technical constraints prevent generation errors that would cause runtime failures during rendering.

The prompt concludes with an explicit JSON output schema specifying the required structure for generated layouts. This schema defines the two-component output format: a suggestion object containing three explanatory text fields (actionable suggestion text, scientific nudge explaining the behavioral rationale, meta-explanation describing why the specific layout was chosen) and a layout array containing widget specification objects with fields for type, title, size, dataSource, and optional chartType. By requesting structured JSON output rather than natural language descriptions of desired layouts, the prompt enables deterministic parsing and direct use of LLM output in the rendering pipeline without additional interpretation or transformation.
All the complete prompts are available in the supplemental material.

\subsection{Widget-Based Rendering Architecture}

The rendering subsystem transforms abstract widget specifications from the layout generation subsystem into concrete interface implementations through a modular pipeline that separates widget instantiation, data binding, visualization rendering, and event handler attachment. This separation of concerns enables clean extension of the widget library and modification of visualization implementations without impacting the adaptive logic.

The rendering pipeline begins with the main renderDashboard function that receives the widget specification array and user context object from the layout generator. This function iterates over specifications, instantiating a DOM container element for each widget with appropriate CSS classes, data attributes for behavioral tracking, and accessibility attributes. Each widget container receives a header region containing the metric title and the explicit feedback controls, ensuring consistent widget chrome across the dashboard. The function examines each widget's type field to dispatch to specialized rendering subfunctions (Table~\ref{tab:widget-types}) that generate the appropriate visualization given the widget's data source specification.

\begin{table*}[t]
\centering
\caption{Widget Visualization Library: Types, Appropriate Data Characteristics, and Rendering Technologies}
\label{tab:widget-types}
\begin{tabular}{@{}llp{5cm}p{4.5cm}@{}}
\toprule
\textbf{Widget Type} & \textbf{Data Type} & \textbf{Optimal Use Case} & \textbf{Implementation} \\
\midrule
\texttt{stat} & Single value & Instantaneous measurements requiring immediate comprehension (steps, weight, heart rate) & Large typography with optional trend arrows, percentage change indicators \\
\addlinespace
\texttt{gauge} & Normalized score & Percentage or 0-100 metrics where position relative to ideal range matters (fitness scores, recovery) & Chart.js doughnut chart configured as semi-circle with color-coded zones \\
\addlinespace
\texttt{chart} & Temporal series & Metrics with meaningful patterns over time requiring trend analysis (weekly steps, sleep) & Chart.js line or bar chart with responsive scaling and hover tooltips \\
\addlinespace
\texttt{heatmap} & Daily binary & Consistency tracking showing presence/absence patterns (workout streaks, goal achievement) & CSS Grid of colored cells with opacity mapped to engagement level \\
\addlinespace
\texttt{timeline} & Event sequence & Chronological activities requiring temporal ordering (meal log, medication schedule) & Vertical list with timestamp and event description pairs \\
\addlinespace
\texttt{stacked\_bar} & Compositional & Part-to-whole relationships where components sum to total (macronutrient breakdown, time allocation) & Chart.js horizontal bar with stacked segments, distinct colors per component \\
\addlinespace
\texttt{radar} & Multi-dimensional & Balanced assessment across multiple related metrics (fitness dimensions, wellness factors) & Chart.js radar chart with normalized axes enabling shape-based comparison \\
\bottomrule
\end{tabular}
\end{table*}

The stat widget renderer implements the simplest visualization modality: a large numerical display with unit label and optional trend indicator. The renderer queries the health data source using the widget's dataSource field, extracts the current value and unit strings, and generates a two-component display with the numerical value rendered in a large typeface and the unit rendered in smaller supplementary text. When temporal trend information exists in the data source, the renderer computes a visual trend indicator showing percentage change with color coding (green for improvements, red for declines relative to health goals) and directional arrows indicating the change direction.

The chart widget renderer implements temporal visualizations using the Chart.js library, supporting line and bar chart modalities for metrics with weekly historical data. The renderer creates a canvas element, binds Chart.js to this canvas, and configures the chart with the metric's temporal values array, label array, and styling preferences. The implementation applies consistent color theming across all charts to maintain visual coherence and configures interaction features including hover tooltips displaying exact values and responsive resizing to handle different widget sizes gracefully.

The gauge widget renderer implements semi-circular gauge visualizations appropriate for percentage or normalized score metrics using Chart.js doughnut charts configured with specific geometric parameters. The renderer computes the visual fill percentage based on whether the metric exhibits positive or negative valence (for example, high heart rate is undesirable while high fitness scores are desirable), applies color coding using a three-tier scheme (red for concerning values, orange for moderate values, green for excellent values), and overlays numerical text in the gauge center to support precise reading. This visualization modality proves particularly effective for communicating normalized health scores that users should understand relative to ideal ranges rather than as absolute measurements.

The heatmap widget renderer implements consistency tracking visualizations showing daily engagement patterns as color-coded grid cells. The renderer creates a grid of div elements, one per day in the tracking period, and sets each cell's background color opacity proportional to the engagement level recorded for that day. This visualization supports immediate perception of behavioral consistency patterns, helping users identify gaps in adherence to health goals without requiring explicit numerical interpretation.

Additional widget renderers implement timeline (chronological event feeds), stacked bar (compositional breakdowns), and radar chart (multi-dimensional balance) visualizations, but the architectural principles remain consistent: each renderer receives an abstract specification with a data source identifier, queries the health data repository, and generates appropriate visualization markup using standard web technologies.

Following widget instantiation and rendering, the system attaches behavioral tracking event handlers to enable the multi-modal feedback channels described earlier. The enableDragAndDrop function attaches HTML5 Drag and Drop API handlers to the container and all widget elements, implementing the spatial reorganization tracking logic. Each widget receives mouseenter and mouseleave handlers implementing attention allocation tracking with the five-hundred millisecond threshold filter. The explicit feedback buttons in each widget header receive click handlers that record preference events and update visual state to provide immediate feedback. This event handler attachment completes the feedback loop, ensuring that subsequent user interactions will inform future layout regenerations.

\section{Implementation and Technical Details}

The released prototype is implemented as a lightweight web application designed to emphasize architectural clarity and reproducibility rather than system scale. A code audit of the submitted repository includes JavaScript engine across seven modules, supplemented by HTML, CSS, a small JSON data store, and package metadata. The implementation targets modern web browsers with ES6 support and employs standard web technologies without framework dependencies. The modular structure separates behavioral tracking (\texttt{drag.js}, \texttt{app.js}), rendering logic (\texttt{render.js}), LLM integration (\texttt{llm.js}), and API management (\texttt{api.js}, \texttt{server.js}) into independent modules that communicate through clearly defined interfaces using JSON as the primary data interchange format.

The frontend implementation is written in vanilla JavaScript using ES module syntax, avoiding framework dependencies to ensure that the architectural approach remains transparent and transferable to alternative implementation contexts. The behavioral tracking modules attach event handlers directly to DOM elements using the standard \texttt{addEventListener} API, maintaining behavioral state in JavaScript \texttt{Map} and \texttt{Set} data structures that support efficient lookup and update operations. The rendering modules generate interface elements through direct DOM manipulation using \texttt{createElement}, \texttt{appendChild}, and related APIs rather than relying on template-based or virtual DOM approaches, ensuring full control over generated markup and simplifying debugging and inspection during development.

The implementation separates four primary runtime concerns. The User Profile Manager maintains questionnaire-derived user attributes such as goals, target values, stress level, preferred visualization style, and desired information density. The Behavioral Tracking Subsystem records per-widget likes and dislikes into persistent history, captures drag-based reordering by updating the stored layout configuration, and accumulates hover dwell times in a session-scoped map. The LLM-Driven Layout Generator assembles these inputs into a structured prompt and requests a JSON-formatted response specifying both a suggestion object and a widget layout. The Rendering Subsystem maps the returned specification to concrete widget instances rendered using browser APIs and Chart.js-based visualizations.

Several implementation characteristics also clarify the scope of the current prototype. Explicit preferences and widget/data compatibility rules are expressed as structured prompt constraints provided to the model, without an additional deterministic post-generation validation layer. Similarly, top-priority widgets are inferred from the leading positions in the persisted layout configuration, providing a practical proxy for user-driven prioritization while remaining sensitive to prior system-generated layouts when no reordering has occurred. These design choices reflect a prompt-constrained and interpretable implementation that prioritizes transparency and modularity, while leaving opportunities for future extensions such as stronger constraint enforcement and richer behavioral inference mechanisms.

The backend implementation provides a minimal Node.js server using the Express framework to serve the frontend application and to proxy requests to the OpenAI API for LLM access. This proxy architecture addresses two technical requirements: it protects the OpenAI API key from exposure in client-side code where it could be extracted and misused, and it enables request logging and rate limiting to prevent API quota exhaustion during development and testing. The custom server is a thin Express proxy of approximately thirty lines wrapping a single layout-generation endpoint; user CRUD is delegated to a json-server instance running on a separate port. User data persists to a JSON file-based database suitable for development and demonstration purposes, though production deployment would substitute a conventional database system.

The layout generation endpoint receives requests containing the full prompt constructed by the frontend, forwards them to the OpenAI API with appropriate authentication headers and model parameters, receives the JSON-formatted response, and returns it to the frontend. The implementation employs the GPT-5-mini model, selected for its balance of generation quality and cost efficiency. The model receives prompts with the system role to establish its task context and operates with default temperature parameter values that provide reasonable diversity in generated layouts without introducing erratic behavior. Error handling logic addresses three common failure modes: OpenAI API unavailability, LLM generation of invalid JSON that fails to parse, and LLM generation of semantically invalid layouts specifying non-existent data sources or widget types. For JSON parsing failures, the implementation attempts basic correction strategies including stripping Markdown code fences that LLMs sometimes emit despite instructions to return raw JSON. For semantic validation failures, the implementation returns an empty layout and logs detailed diagnostics to support prompt engineering refinement.

The health data repository implements a simulated personal health monitoring data source providing realistic measurements across fourteen instantaneous metrics and a set of temporal and structured data sources (Table~\ref{tab:health-metrics}). The prototype uses synthetic but plausible health-style data to enable demonstration of the architectural behavior without requiring integration with live device APIs.
In the current implementation, the repository exposes fourteen instantaneous metrics through a \texttt{user\_stats} structure and a set of richer data sources used for visualization, including weekly trends, daily traces, compositional breakdowns, a recent-activity timeline, and a heatmap-style consistency view. This arrangement is sufficient to exercise the widget library and to evaluate whether the layout generator adapts modality, prominence, and information density in response to user profile and interaction signals.
The application domain is used here as a technically demanding setting for adaptive dashboard composition rather than as a source of clinically meaningful health inference.

\begin{table*}[t]
\centering
\caption{Health Metrics Available in the System Data Repository}
\label{tab:health-metrics}
\begin{tabular}{@{}llp{6cm}l@{}}
\toprule
\textbf{Metric Identifier} & \textbf{Category} & \textbf{Description} & \textbf{Unit} \\
\midrule
\multicolumn{4}{l}{\textit{Instantaneous Measurements (for stat/gauge widgets)}} \\
\midrule
today\_steps & Activity & Daily step count accumulated & steps \\
today\_resting\_hr & Cardiovascular & Resting heart rate baseline & bpm \\
today\_walking\_hr & Cardiovascular & Average heart rate during walking & bpm \\
today\_hr\_variability & Recovery & Heart rate variability indicating autonomic balance & ms \\
today\_exercise\_minutes & Activity & Total active exercise time & minutes \\
today\_time\_sleep & Recovery & Sleep duration from previous night & hours \\
today\_climbed\_floor & Activity & Floors climbed via stairs & floors \\
today\_breath\_frequency & Respiratory & Breathing rate measurement & breaths/min \\
walking\_speed & Mobility & Average walking pace & km/h \\
stair\_speed & Mobility & Stair climbing velocity & floors/min \\
step\_length & Gait & Average stride length & cm \\
today\_asymmetric\_walk & Gait & Gait symmetry score (lower is better) & score \\
today\_stand\_up\_time & Activity & Hours spent standing & hours \\
vo2\_max & Fitness & Estimated cardiorespiratory fitness & mL/kg/min \\
\midrule
\multicolumn{4}{l}{\textit{Temporal Series (for chart widgets)}} \\
\midrule
steps\_weekly & Activity & Seven-day step count history & steps/day \\
sleep\_weekly & Recovery & Seven-day sleep duration history & hours/night \\
heart\_rate\_weekly & Cardiovascular & Seven-day resting heart rate trend & bpm \\
calories\_weekly & Nutrition & Seven-day caloric intake and expenditure & kcal/day \\
weight\_weekly & Body Composition & Seven-day weight measurements & kg \\
exercise\_weekly & Activity & Seven-day exercise time history & minutes/day \\
water\_weekly & Hydration & Seven-day water intake tracking & glasses/day \\
nutrition\_weekly & Nutrition & Seven-day macronutrient composition & grams/day \\
\bottomrule
\end{tabular}
\end{table*}

The instant metrics include step count, resting heart rate, walking heart rate, heart rate variability, active exercise minutes, sleep duration, floors climbed, breathing frequency, walking speed, stair climbing speed, step length, gait asymmetry score, standing hours, and VO2 max cardiorespiratory fitness estimate. The temporal series provide weekly historical values for steps, sleep, calories, weight, heart rate, exercise minutes, water intake, and nutrition composition. The simulation generates values with realistic ranges and temporal patterns including weekday versus weekend variation, enabling demonstration of system capabilities without requiring integration with actual health monitoring APIs. The data structure uses a hierarchical JSON format with separate objects for instant measurements and chart series, each containing value, unit, timestamp, and styling metadata enabling visualization without additional configuration.

The implementation employs Chart.js as the visualization library, selected for its comprehensive chart type support, responsive design capabilities, and extensive customization options enabling consistent visual theming. The widget rendering modules configure Chart.js instances with global defaults including color schemes matching the dashboard aesthetic, interaction modes enabling hover tooltips and click events for drill-down where appropriate, and responsive sizing that adapts visualizations to different widget dimensions. Custom styling ensures visual consistency across widget types, with all visualizations employing the same accent colors, typography, and spacing parameters.

The CSS implementation employs CSS Grid layout for the dashboard structure, enabling flexible two-dimensional widget positioning with automatic responsiveness to different screen sizes. The grid configuration defines a repeating column structure with a minimum column width constraint ensuring readability on small screens while supporting multiple columns on larger displays. Widgets specify their grid span through CSS classes (normal for single-column widgets, wide for widgets spanning two columns), with the grid layout engine automatically handling wrapping and alignment. This approach provides the flexibility required for LLM-generated layouts specifying arbitrary widget arrangements while maintaining responsive behavior without media query complexity.

\section{System Demonstration and Analytical Evaluation}

The following section presents an evaluation of the proposed architecture using data collected from six users who interacted with the system across one or more sessions.
Rather than relying on synthetic behavioral traces, the evaluation is grounded in profiles and interaction histories recorded directly by the system during use.
The goal is to examine whether the architecture produces consistent, interpretable, and goal-aligned adaptations across users with meaningfully different health objectives, interface preferences, and engagement styles.

\subsection{User Profiles}

Six users interacted with the system, each completing the onboarding questionnaire before receiving an AI-generated dashboard.
The questionnaire collected self-reported demographic information, health goals, current and target weight, preferred step and sleep targets, and a set of interface preferences covering information density, visual aesthetic, and visualization style.
Table~\ref{tab:user_profiles} summarizes the attributes recorded at registration.

The group covers a wide age range, from 22 to 67 years, and comprises three women and three men.
Primary health goals span five distinct objectives: weight reduction (Sarah and Marcus), increasing general daily activity (James, who framed this as wanting to ``move a bit more''), maintaining and supporting athletic performance (Elena), scheduling structured exercise within a demanding routine (Priya), and improving sleep quality (Tyler).
This diversity in stated goals is directly consequential for the adaptation architecture: the layout generation component must reason differently about each user from the very first session.
A weight-loss profile calls for weight trend and calorie tracking widgets; a performance profile foregrounds cardiorespiratory metrics such as VO2 max and heart rate variability; a sleep-focused profile surfaces sleep duration and stage breakdown indicators.
The profile section of the structured prompt, described in Section~3.3, translates these goal differences into distinct design priorities that persist across regeneration cycles.

Activity levels span the full range of the questionnaire options.
Elena reports competitive athletic training, Sarah and Priya report moderate habitual activity, and both Marcus and James identify as sedentary, a detail that is particularly relevant for the two users whose stated goal involves increasing movement.
Stress levels also vary across the group: Priya reports high stress, Marcus and Tyler report medium stress, and the remaining three report low stress.
These attributes influence how the temporal prompt layer frames adaptation priorities, for instance by promoting stress-related widgets and recovery indicators for users with elevated stress scores.

Interface preferences are distributed across multiple dimensions.
Three users selected simple information density (Sarah, Elena, and Tyler) and three selected detailed (James, Marcus, and Priya).
The aesthetic mood preferences cover all three available styles: professional for Sarah and Elena, friendly for James and Marcus, and gamified for Priya and Tyler.
Five of the six users expressed a preference for chart-based visualizations, while Elena preferred compact statistical displays, a choice that led the system to generate a stat-heavy layout even for metrics that could otherwise be rendered as trend lines.
Two users, James and Tyler, indicated that they wear glasses, a signal that is included in the user-profile section of the structured prompt, allowing the LLM to favour larger widgets and lower information density.

Regarding interaction history, three users generated behavioral signals during their session and three did not.
Sarah disliked two widgets, specifically today's step count and resting heart rate, which the system interpreted as hard exclusion constraints for subsequent regenerations.
Marcus disliked one widget (the sleep quality breakdown), and Elena disliked one widget (VO2 max) while also producing spatial reorganization events and sustained hover signals that the dwell-time tracking subsystem recorded.
James and Priya generated only spatial reorganization and dwell signals, without using the explicit feedback affordances. Tyler completed onboarding without generating any interaction signals, providing a pure profile-driven baseline.

\begin{table*}[t]
  \caption{
    Profile attributes of the six users collected through the onboarding questionnaire.
    \textit{Activity} values correspond to the questionnaire options: Sedentary, Moderate, and Athlete.
    \textit{Complexity} refers to preferred information density (Simple or Detailed).
    \textit{Vis. Pref.} denotes the preferred visualization style (Charts or Stats).
    \textit{Behavioral Signals} summarizes the interaction history recorded by the system during the session;
    ``dislike'' entries are \texttt{widget\_preference} events with negative sentiment;
    ``drag'' and ``dwell'' refer to spatial reorganization and hover-time events respectively.
  }
  
  \label{tab:user_profiles}
  \footnotesize

  \begin{tabularx}{\textwidth}{@{} l l l p{2.2cm} l l l l l X @{}}
    \toprule
    \textbf{User}  &
    \textbf{Age}   &
    \textbf{Gender} &
    \textbf{Primary Goal} &
    \textbf{Activity} &
    \textbf{Stress} &
    \textbf{Complexity} &
    \textbf{UI Mood} &
    \textbf{Vis. Pref.} &
    \textbf{Behavioral Signals} \\
    \midrule
    Sarah  & 31 & F & Weight reduction              & Moderate  & Low    & Simple   & Professional & Charts & 2 widget dislikes (step count, resting HR)       \\[2pt]
    Elena  & 26 & F & Maintain athletic performance & Athlete   & Low    & Simple   & Professional & Stats  & 1 widget dislike (VO2 max); drag reorder; dwell  \\[2pt]
    James  & 67 & M & Increase daily activity       & Sedentary & Low    & Detailed & Friendly     & Charts & drag reorder; dwell                             \\[2pt]
    Marcus & 45 & M & Weight reduction              & Sedentary & Medium & Detailed & Friendly     & Charts & 1 widget dislike (sleep quality breakdown)       \\[2pt]
    Priya  & 33 & F & Schedule regular exercise     & Moderate  & High   & Detailed & Gamified     & Charts & drag reorder; dwell                              \\[2pt]
    Tyler  & 22 & M & Improve sleep quality         & Moderate  & Medium & Simple   & Gamified     & Charts & None                              \\
    \bottomrule
  \end{tabularx}
\end{table*}
Taken together, the six profiles are sufficiently diverse to exercise several distinct branches of the adaptation logic simultaneously.
The contrast between user with no interaction history and those with rich behavioral signals allows the evaluation to isolate the contribution of each adaptation layer: profile-driven cold-start generation for Tyler; explicit feedback enforcement for Sarah and Marcus; spatial reorganization and attention allocation for James and Priya and full multi-signal synthesis combining feedback, spatial priority, and attention allocation for Elena.
The overlap between Sarah and Marcus, both targeting weight reduction but with different activity levels, stress profiles, and aesthetic preferences, additionally allows an examination of how profile-level differences produce divergent layouts even when the stated goal is identical.

\subsection{System Walkthrough}

Before examining per-user adaptation outcomes, this section provides a concrete walkthrough of the three-stage interaction pipeline as experienced by a representative user.
The walkthrough is illustrated using screenshots from Sarah's session and covers onboarding through profile creation, initial dashboard generation, and on-demand health report generation.
Screenshots from all six participants are available in the supplemental material.

\begin{figure}[t]
  \centering
  \includegraphics[width=0.6\columnwidth]{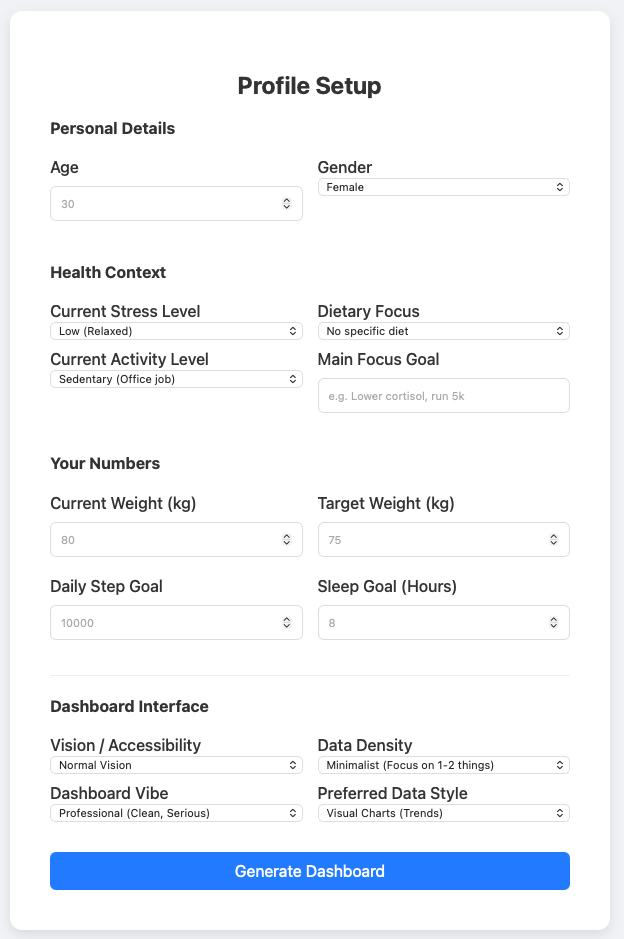}
  \caption{%
    Profile Setup form completed by Sarah.
    The questionnaire is organised into four sections covering personal details, health context, quantitative targets, and interface preferences.
    Each field maps to a specific section of the structured prompt used by the layout generation subsystem.
    The \textit{Dashboard Interface} section is of particular relevance: the selected data density (Minimalist), dashboard vibe (Professional), and preferred data style (Visual Charts) directly constrain the widget types and layout density produced by the AI.
  }
  \label{fig:profile}
  \Description{Profile Setup form completed by Sarah.}
\end{figure}

\paragraph{Profile creation.}
The entry point to the system is a structured onboarding form organised into four sections (Figure~\ref{fig:profile}).
\textit{Personal Details} collects age and gender.
\textit{Health Context} captures current stress level, dietary focus, habitual activity level, and a free-text field for the primary health goal, which allows users to express objectives in natural language rather than selecting from a fixed list.
\textit{Your Numbers} records current and target weight alongside daily step and sleep goals, providing the system with quantitative targets that anchor goal-aware reasoning during layout generation.
\textit{Dashboard Interface} collects four interface preference attributes: visual accessibility needs, preferred information density, aesthetic mood (ranging from professional and clean to gamified), and preferred data representation style (visual charts versus compact statistics).
These four sections map directly to the user profile section of the structured prompt described in Section~3.3, ensuring that every questionnaire field has an explicit role in the adaptation logic.
Once the form is submitted, the profile is persisted to the user database and the layout generation pipeline is invoked immediately.

\begin{figure*}[t]
  \centering
  \includegraphics[width=0.92\textwidth]{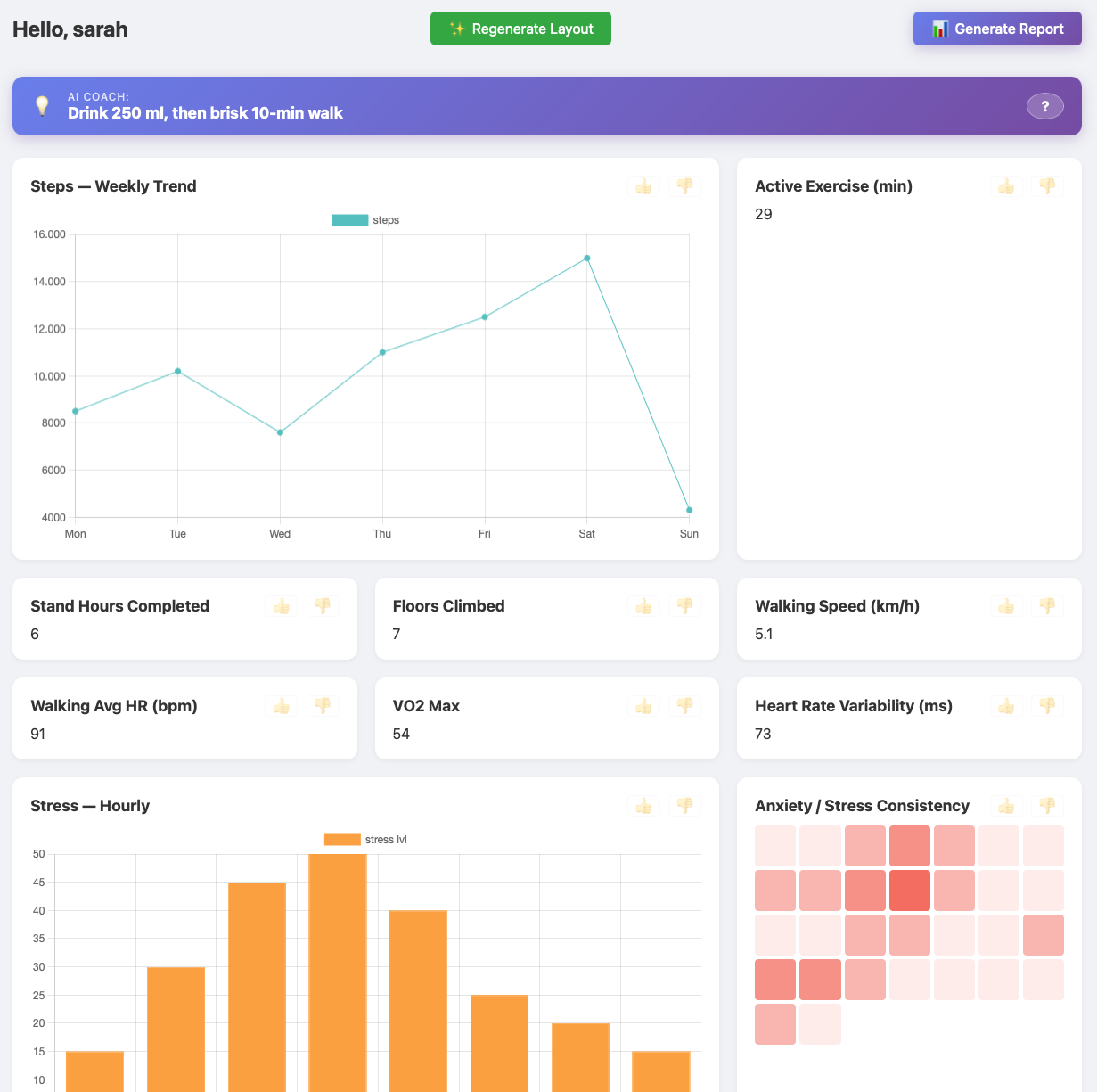}
  \caption{%
    Adaptive dashboard generated for Sarah.
    The AI Coach banner at the top presents a contextually appropriate action suggestion alongside a toggle for the scientific and meta-level explanation.
    The widget grid combines a dominant weekly steps trend chart (wide, reflecting her chart preference and weight-loss goal), six compact activity and physiological stat widgets, a wide hourly stress chart, and an anxiety heatmap.
    Today's step count and resting heart rate are absent from the layout: both were explicitly disliked by the user and were excluded as hard constraints by the prompt engineering layer.
    Per-widget like and dislike affordances (visible in each widget header) allow the user to continue shaping the layout across sessions.
  }
  \label{fig:dashboard}
  \Description{Adaptive dashboard generated for Sarah.}
\end{figure*}

\paragraph{Dashboard generation.}
Figure~\ref{fig:dashboard} shows the dashboard generated for Sarah following her initial interaction session.
The interface is structured around three visual layers.
At the top, the AI Coach banner displays an actionable suggestion derived from the suggestion object returned by the layout generator: in this case, ``Drink 250\,ml, then brisk 10\,min walk,'' accompanied by a \textit{?} button that expands a pop-up containing the scientific rationale and a meta-explanation of the layout choices.
Below the banner, the main grid presents a mixed layout of chart and statistical widgets.
The first and most prominent widget is a wide line chart showing the weekly step trend, consistent with Sarah's stated goal of weight reduction and her preference for visual chart representations.
Compact stat widgets follow for activity-related metrics including exercise minutes, stand hours, floors climbed, walking speed, walking average heart rate, VO2 max, and heart rate variability.
A wide bar chart showing hourly stress levels and a heatmap of anxiety consistency patterns occupy the lower portion of the visible grid.

Notably, neither a today's step count widget nor a resting heart rate widget appears anywhere in the layout, despite both being available in the data repository and relevant to a weight-loss profile.
This absence directly reflects Sarah's two explicit dislikes recorded in her interaction history: the system treated both signals as hard exclusion constraints during prompt construction and the generated layout respected them unconditionally.
Each widget carries a pair of lightweight like and dislike affordances in its header, enabling Sarah to continue refining her preferences in subsequent sessions.
The \textit{Regenerate Layout} and \textit{Generate Report} buttons in the header provide access to on-demand adaptation and the analytical report feature respectively.

\begin{figure*}[t]
  \centering
  \begin{subfigure}[t]{0.48\textwidth}
    \centering
    \includegraphics[width=\linewidth]{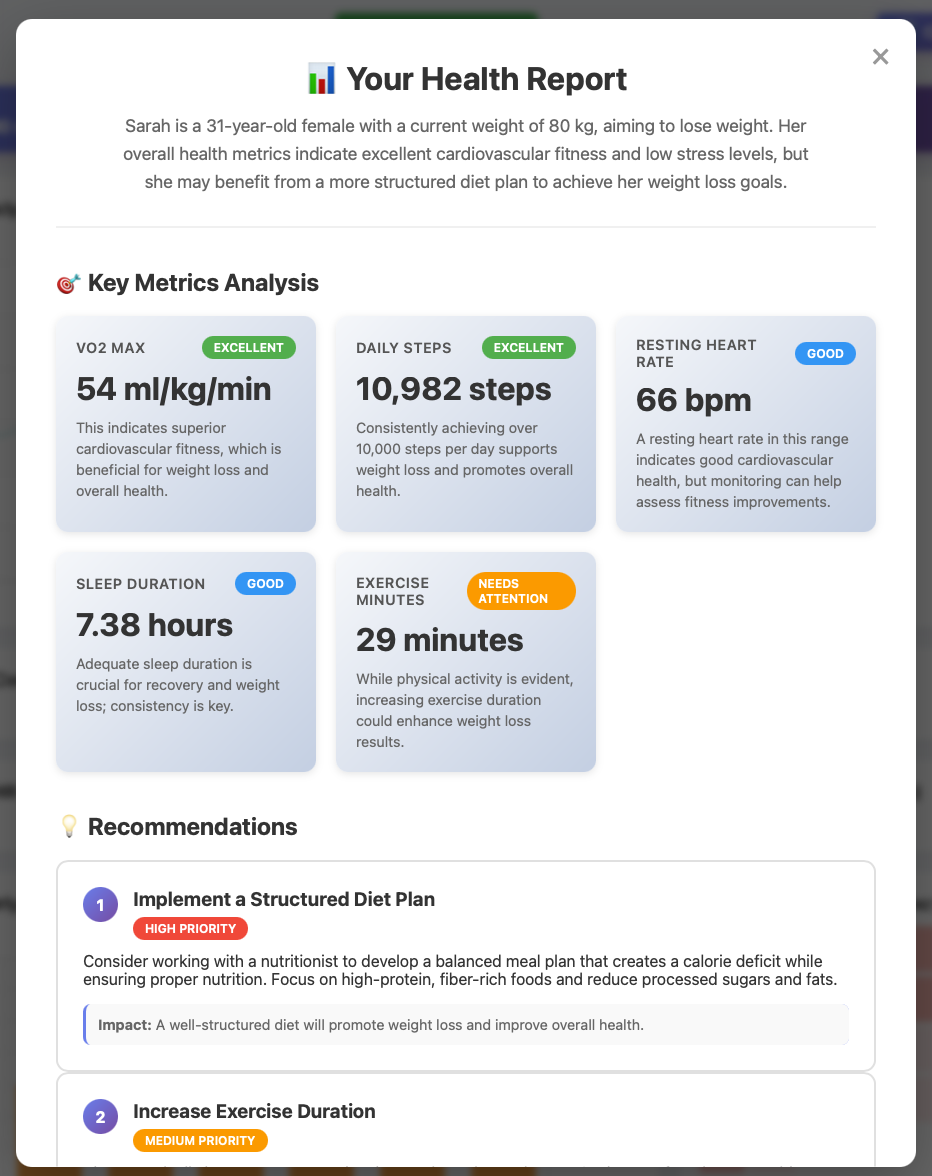}
    \caption{%
      Upper section of the health report showing the personalised summary and the Key Metrics Analysis cards.
      Each card carries a qualitative status badge and a goal-contextualised interpretation.
      Exercise duration is flagged as \textit{Needs Attention}, directly informing the high-priority recommendation in the section below.
    }
    \label{fig:report-top}
      \Description{health report}

  \end{subfigure}
  \hfill
  \begin{subfigure}[t]{0.48\textwidth}
    \centering
    \includegraphics[width=\linewidth]{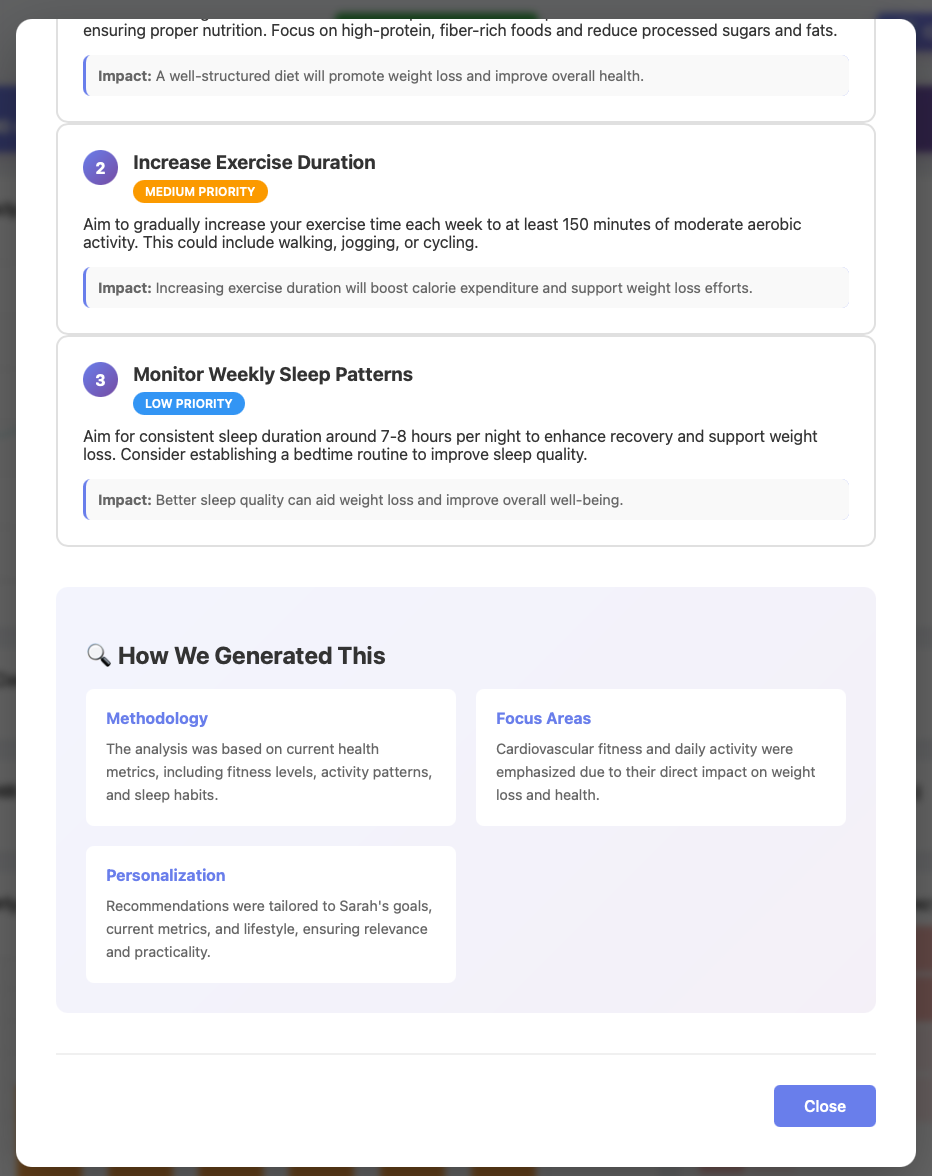}
    \caption{%
      Lower section of the health report showing the prioritised recommendations and the explainability panel.
      The three \textit{How We Generated This} cards expose the analytical methodology, focus area selection rationale, and personalisation logic, providing transparency at the meta-reasoning level.
    }
    \label{fig:report-bottom}
          \Description{health report}

  \end{subfigure}
  \caption{%
    Health report generated for Sarah on demand.
    The report is structured into three components: a plain-language summary, a key metrics analysis with status-rated cards, and a prioritised recommendations list each accompanied by an impact statement.
    A dedicated explainability section closes the report by articulating the reasoning behind both the analytical focus and the personalisation decisions.
    The full report is produced in a single model inference using the same structured prompt framework as the layout generator.
  }
  \label{fig:report}
        \Description{health report}

\end{figure*}

\paragraph{Health report generation.}
On request, the system generates a personalised health report rendered as a modal overlay (Figure~\ref{fig:report}).
The report is structured around three distinct components that parallel the explainability design described in Section~3.3.
The opening summary provides a two-to-three sentence synthesis of the user's overall health status written in plain language and anchored to their stated goal.
For Sarah, the report identifies strong cardiovascular fitness (VO2 max of 54\,ml/kg/min, rated \textit{Excellent}) and healthy daily activity (10,982 steps, rated \textit{Excellent}), while flagging exercise duration (29 minutes, rated \textit{Needs Attention}) as the primary improvement target given her weight-loss objective.
The \textit{Key Metrics Analysis} section then presents each assessed metric as an individual card carrying a label, a current value, a qualitative status badge, and a one-sentence interpretation grounding the number in the user's goal context.

The \textit{Recommendations} section lists prioritised action items, each with a title, a priority badge (high, medium, or low), a detailed description, and an explicit impact statement.
In Sarah's report, three recommendations were generated: implementing a structured dietary plan (high priority), increasing weekly exercise duration toward 150 minutes of moderate aerobic activity (medium priority), and monitoring sleep patterns to support recovery and weight management (low priority).
The \textit{How We Generated This} section closes the report with three explainability cards describing the analytical methodology, the focus areas selected and why, and the personalisation logic applied.
This three-tier structure, covering what to do, why it was recommended, and how the reasoning was constructed, supports transparency at multiple levels and is generated as part of a single model inference rather than through a separate post-hoc explanation pipeline.

\subsection{Observed Interaction Patterns}

Interaction signals recorded across the six sessions were heterogeneous both in type and in quantity.
Table~\ref{tab:signals} summarises which behavioral channels were active for each user.
Five of the six users generated at least one signal that the adaptation pipeline could act on, while Tyler completed onboarding and received a layout without any further interaction, providing a pure profile-driven baseline.
The distribution covers all three tracking channels in isolation and in combination, allowing the evaluation in the following subsection to compare the contribution of each layer independently.

\begin{table}[t]
  \caption{%
    Behavioral signals recorded per user.
    Explicit feedback refers to widget-level like or dislike events persisted to the interaction history.
    Spatial reorganization refers to drag-and-drop reordering events that updated the stored layout.
    Dwell time refers to hover events exceeding the 500\,ms threshold recorded by the attention allocation channel.
    The total active channels column counts the distinct signal types present for each user.
  }
  \label{tab:signals}
  \footnotesize
  \begin{tabular}{lcccc}
    \toprule
    \textbf{User} &
    \textbf{Explicit} &
    \textbf{Spatial} &
    \textbf{Dwell} &
    \textbf{Active} \\
    & \textbf{Feedback} & \textbf{Reorder} & \textbf{Time} & \textbf{Channels} \\
    \midrule
    Sarah  & \checkmark (2 dislikes) & --         & --         & 1 \\
    Elena  & \checkmark (1 dislike)  & \checkmark & \checkmark & 3 \\
    James  & --                      & \checkmark & \checkmark & 2 \\
    Marcus & \checkmark (1 dislike)  & \checkmark & \checkmark & 3 \\
    Priya  & --                      & \checkmark & \checkmark & 2 \\
    Tyler  & --                      & --         & --         & 0 \\
    \bottomrule
  \end{tabular}
\end{table}

\paragraph{Explicit feedback.}
Three users interacted with the like and dislike affordances embedded in widget headers.
Sarah used the dislike button on two widgets: today's step count and resting heart rate.
Both dismissals are consistent with her stated preference for chart-based trend visualisations rather than single-value displays: a raw step count and a standalone heart rate reading offer less contextual information than the weekly trend charts.
Elena disliked the VO2 max widget, a signal that stands in notable contrast with her athletic profile, for which VO2 max is ordinarily a primary performance indicator.
One plausible interpretation is that Elena considers this metric too static for daily monitoring, preferring real-time activity signals such as exercise minutes and heart rate variability.
Marcus disliked the sleep quality breakdown widget, which is consistent with his interaction history suggesting a focus on activity and cardiovascular indicators rather than sleep architecture detail.
In all three cases, the system recorded the dislike as a hard exclusion constraint, and the subsequent layout generation respected this unconditionally regardless of goal-profile relevance.

\paragraph{Spatial reorganization.}
Four users performed drag-and-drop reordering at some point during their session.
Elena dragged exercise minutes, heart rate variability, and hourly stress to the top three positions of the grid, establishing an implicit priority signal oriented toward acute physiological and psychological readiness rather than cumulative statistics.
Marcus reorganised his layout so that floors climbed, VO2 max, and stand hours occupied the leading positions, a configuration that differs from the AI-generated initial ordering and reflects his personal weighting of effort-based movement indicators over aggregate step counts.
James and Priya both performed reordering consistent with their respective goals: James promoted step-count and active-minutes widgets toward the top of the grid, aligning with his objective of increasing general daily movement; Priya elevated step count and exercise minutes alongside stress-related widgets, reflecting both her activity scheduling goal and her high reported stress level.
In all four cases, the persisted layout order was updated in the database immediately following the drag event, making the spatial signal available to the next regeneration cycle.
The system interprets the top three positions of the stored layout as the current user-defined priority set, a heuristic that translates the drag action into a concrete design instruction without requiring any additional user input.

\paragraph{Dwell time.}
The attention allocation channel recorded sustained hover events for Elena, James, Marcus, and Priya.
Elena accumulated significant dwell time on the weekly steps chart, the recent activities timeline, and the anxiety heatmap; all three were subsequently rendered at wide size in the regenerated layout, consistent with the system interpretation of prolonged attention as a signal warranting expanded detail.
Marcus dwelled most on the anxiety heatmap and the resting heart rate widget, both of which appear at wide size in his stored layout, corroborating the prompt-level instruction to expand highly studied widgets.
For James, dwell concentrated on cardiovascular indicators, particularly resting heart rate and the daily heart rate trace, a pattern interpretable as health-monitoring vigilance consistent with his age.
Priya's most sustained hover activity occurred on the hourly stress chart and the anxiety heatmap, which aligns with her high self-reported stress level and suggests an investigative engagement with stress pattern data rather than incidental cursor movement.
As described in Section~3.2, dwell time is treated as a soft signal: the system interprets it as a preference for additional detail but does not treat it as a hard constraint in the way that explicit dislikes are handled.
The resulting adaptation is therefore a suggestion to widen or deepen the presentation of studied widgets rather than a guaranteed inclusion rule.

\paragraph{Tyler as a cold-start baseline.}
Tyler completed the onboarding questionnaire and received a dashboard without generating any interaction signals during the session.
His layout is therefore entirely determined by profile-driven reasoning: the system used his stated goal of improving sleep quality, his medium stress level, his moderate activity, and his preference for a simple gamified interface to select and order widgets without any behavioral evidence to build on.
This case represents the system operating in its most constrained mode, relying exclusively on questionnaire data and temporal context, and it provides a clean reference point for assessing the additional contribution of behavioral signals in the five other cases.

\subsection{Per-User Adaptation Narratives}

This subsection presents the adaptation outcome for each user, connecting the behavioral signals described in Section~5.3 to the layout decisions produced by the generation pipeline.
Because only post-adaptation dashboard states are available, the analysis does not compare before and after layouts directly; instead, it reasons from the stored layout structure, the persisted suggestion and explainability objects, and the interaction signals to reconstruct the adaptation logic.
Dashboard screenshots for all six users are provided in the supplemental material.

\paragraph{Sarah.}
Sarah's health data reflects a moderately active profile with solid cardiorespiratory indicators (VO2 max 54\,ml/kg/min, resting HR 66\,bpm, HRV 73\,ms) and adequate but not surplus sleep (7.38\,hrs), with exercise duration (29\,min) identified by the report as the primary improvement target relative to her weight-loss goal.
Her two explicit dislikes, today's step count and resting heart rate, were the only behavioral signals the system had to work with, and both are entirely absent from the final layout regardless of their relevance to her goal profile.
The adaptation instead surfaces these same metrics through temporal and trend representations: the weekly steps trend appears as the first and widest widget, and cardiovascular data is embedded within the activity stat cluster rather than as a standalone HR reading.
The system's explainability identifies the governing logic as a combination of goal-driven inclusion (weight trend, exercise minutes), temporal context (active daytime phase favouring actionable data), and hard-constraint enforcement on the two dismissed widgets.
The outcome illustrates a case where explicit feedback alone is sufficient to redirect the presentation modality of a metric without eliminating the underlying information.

\paragraph{Elena.}
Elena's metrics reflect an elite training profile: VO2 max 68\,ml/kg/min, resting HR 44\,bpm, HRV 112\,ms, 94 exercise minutes, and 18,430 steps on the recorded day, with sleep consistently above 7.8\,hours across the weekly trace.
All three behavioral channels were active, making her the most signal-rich case in the corpus.
The drag reorder placed exercise minutes, HRV, and hourly stress at the top three positions, establishing a priority frame oriented toward readiness and acute load rather than cumulative output.
Her dwell time on the weekly steps chart, the recent activities timeline, and the anxiety heatmap caused all three to be rendered at wide size in the regenerated layout, expanding their visual real estate to match her investigative engagement.
The dislike of VO2 max is the most counterintuitive signal in the dataset: for an athlete, this is typically a primary metric, yet Elena's dismissal was treated as a hard exclusion.
One interpretation, consistent with her preference for compact stats over charts, is that a single static number carries less actionable value per session than dynamic signals such as HRV and exercise duration.
The explainability object confirms that the system synthesised all three signal types simultaneously, preserving the dragged priority order at the top, widening the dwelled widgets, and removing VO2 max without substitution.

\paragraph{James.}
James is the oldest user in the group (67 years) with a fully sedentary baseline: VO2 max 29\,ml/kg/min, resting HR 71\,bpm, HRV 28\,ms, and 7,215 steps on the recorded day, against a weight target of 80\,kg from a current 85\,kg.
His session produced no explicit feedback, making drag reorder and dwell time the sole signals available to the pipeline.
The resulting layout places daily step count at the top as a wide stat followed immediately by the seven-day steps trend, a configuration consistent with his drag-promoted positioning of activity metrics and well matched to his stated goal of moving more.
His dwell activity concentrated on cardiovascular content, particularly resting heart rate and the daily heart rate trace; both appear in the layout but do not occupy leading positions, reflecting the system's interpretation of dwell as a signal for expanded detail rather than spatial priority.
The friendly UI mood and detailed complexity preference are visible in the layout organisation: a sequence of large, readable stat and gauge widgets precedes the chart content, and the recent activities timeline appears at full width at the bottom, providing a narrative account of the day that suits a reflective browsing style.

\paragraph{Marcus.}
Marcus presents the most behaviorally dense case after Elena: a sedentary, medium-stress office worker (VO2 max 38\,ml/kg/min, resting HR 78\,bpm, HRV 34\,ms, 5,421 steps, only 8 exercise minutes, and a weight of approximately 85\,kg trending slowly downward) who activated all three signal channels.
His explicit dislike of the sleep quality breakdown was enforced as a hard exclusion; the widget does not appear in the layout despite sleep duration being a relevant metric for a user with only 6.1\,hours recorded.
His dwell activity on the anxiety heatmap and resting heart rate caused both to be rendered at wide size, a notable outcome given that his resting HR of 78\,bpm sits at the upper boundary of the normal range and his anxiety heatmap shows the highest intensity values in the dataset, tied with Priya.
The drag reorder signal, however, reveals a particularly informative divergence: the explainability object records that the AI generation prioritised daily steps, exercise minutes, and heart rate over the full day at the top of the layout, yet the current stored layout shows floors climbed, VO2 max, and stand hours in those positions.
This discrepancy indicates that Marcus performed a post-generation drag reorder, repositioning his preferred metrics after inspecting the AI-generated result.
The stored order now represents his personal priority override, and it will serve as the spatial signal input for the next regeneration cycle, meaning the adaptation loop will treat effort-based movement indicators as his current top priorities rather than aggregate step counts.

\paragraph{Priya.}
Priya's data captures the profile of a high-stress user with moderate but inconsistent activity (VO2 max 46\,ml/kg/min, resting HR 69\,bpm, HRV 58\,ms, 9,340 steps, 35 exercise minutes, sleep 6.25\,hrs) and a stress trace with values reaching 72 at peak afternoon hours.
Her interaction relied on drag reorder and dwell time without any explicit feedback events, and her goal of fitting exercise into a busy schedule drove the profile-level framing of the layout.
Steps and exercise minutes occupy the first two positions as wide and normal stats respectively, consistent with her drag-promoted activity emphasis, while the HRV gauge immediately follows as a proxy for real-time stress readiness.
Her dwell activity on the hourly stress chart and the anxiety heatmap, both of which are present in the mid-section of the layout, reflects the investigative engagement pattern described in Section~3.2: the system widened or retained these widgets to satisfy her apparent need for stress pattern detail without promoting them above her dragged activity priority.
The gamified UI mood is expressed through the layout's use of goal-oriented stat widgets and progress-style gauges at the top rather than the chart-heavy opening that a professional vibe would have produced, and the radar chart for gait and pace appears as a secondary detail consistent with her detailed complexity preference.

\paragraph{Tyler.}
Tyler's health data shows a cardiovascularly capable but sleep-deprived profile: VO2 max 49\,ml/kg/min, resting HR 62\,bpm, HRV 82\,ms, and 55 exercise minutes, alongside a sleep duration of 5.9\,hours and a weekly sleep trace that oscillates from 3.8 to 10.0\,hours, exposing pronounced irregularity rather than a chronic deficit.
With no interaction signals of any kind, the layout is generated exclusively from his questionnaire profile and the temporal context at the time of the session.
The system placed sleep metrics in the first half of the layout: last night's sleep duration appears as an early stat widget, the seven-day sleep trend occupies a wide chart position, and the sleep quality breakdown is included as a stacked bar, all driven by his stated goal of improving sleep.
Activity and stress content accompanies the sleep cluster, consistent with the daytime phase instruction to balance actionable monitoring with goal-relevant tracking.
The explainability object for Tyler is consequently the most transparent in the dataset: in the absence of behavioral evidence, it can articulate only profile-level and temporal reasoning, offering no reference to user-driven signals and relying entirely on the questionnaire data and the time-of-day heuristic.

\subsection{Health Report Generation}

The health report is an on-demand feature invoked independently of the adaptive dashboard: it uses the same user profile and health data repository but issues a separate inference request to the language model, producing a structured document rather than a widget layout specification.
Its output comprises four components that mirror the explainability design described in Section~3.3: a two-to-three sentence plain-language summary, a key metrics analysis section with status-rated cards, a prioritised recommendations list, and a three-card explainability panel.
While the dashboard adapts continuously across sessions, the report provides a snapshot interpretation of the full data at a single point in time, making it complementary rather than redundant.
Reports for all six users are provided in the supplemental material (Figures~S7--S12); this subsection focuses on what the cross-user variation reveals about the report module's reasoning behaviour.

\paragraph{Structural variation across users.}
The report structure is consistent across all users, but the content varies in three analytically meaningful ways.
First, the set of metrics surfaced in the key metrics section is not fixed: the model selects which indicators to analyse based on the profile and goal, and the number of cards ranges from four (James and Tyler) to five (Sarah, Elena, and Priya).
Second, status badges differ both in label and in semantic weight: all of Elena's metrics receive either \textit{Excellent} or \textit{Good}, while Priya is the only user whose report includes a \textit{Concerning} badge, assigned to her self-reported stress level.
Third, the number and priority distribution of recommendations varies: Tyler receives the fewest (two), while Priya is the only user with two simultaneous high-priority flags.
These differences are not artefacts of the output format but reflect the model performing differentiated reasoning about each profile, which is the expected behaviour of a goal-sensitive report generator.

\begin{figure*}[t]
  \centering
  \begin{subfigure}[t]{0.48\textwidth}
    \centering
    \includegraphics[width=\linewidth]{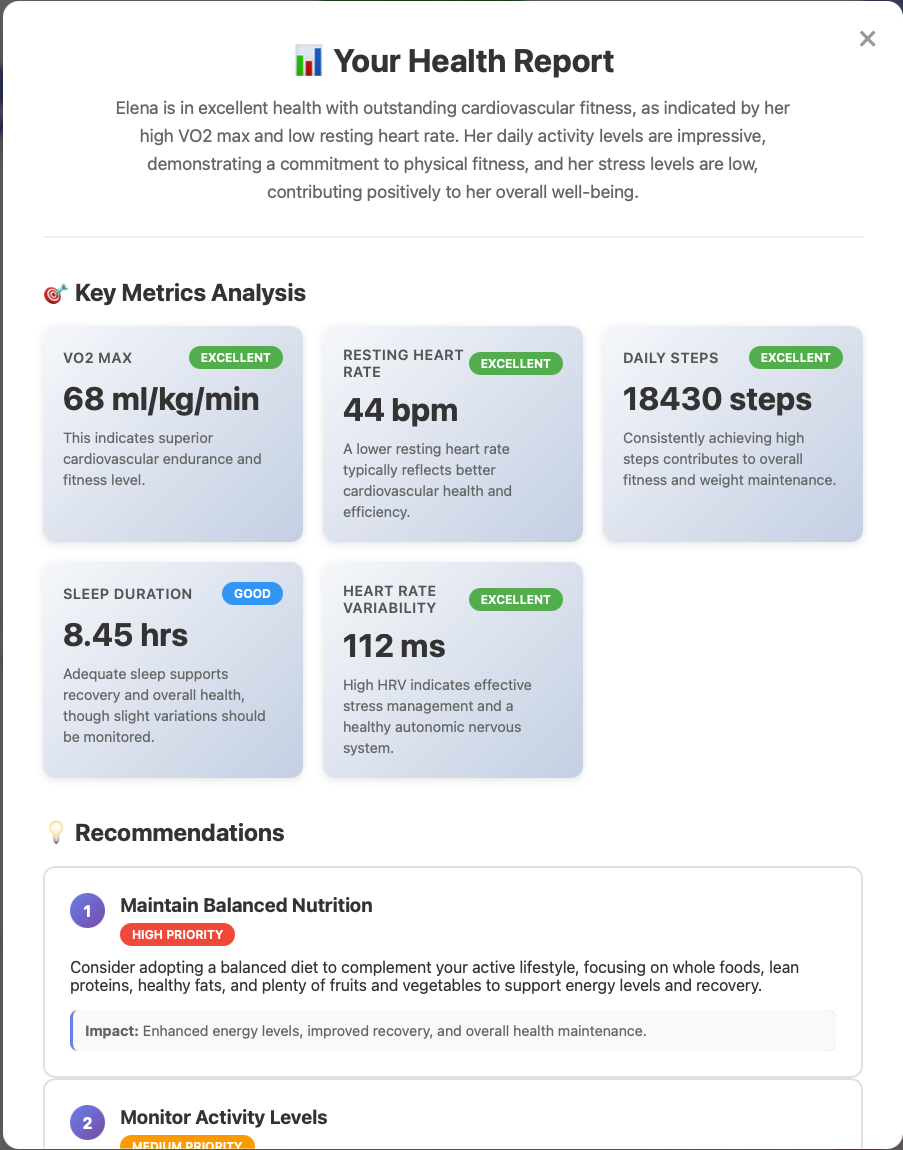}
    \caption{%
     Upper section of Elena's health report.
      All five assessed metrics receive Excellent or Good status badges.
      The absence of threshold violations does not prevent the model from generating substantive recommendations; instead it shifts focus to nutritional support and overtraining prevention, demonstrating goal-sensitive reasoning beyond reactive flagging.
    }
    \label{fig:elena-report-top}
          \Description{Elena report}

  \end{subfigure}
  \hfill
  \begin{subfigure}[t]{0.48\textwidth}
    \centering
    \includegraphics[width=\linewidth]{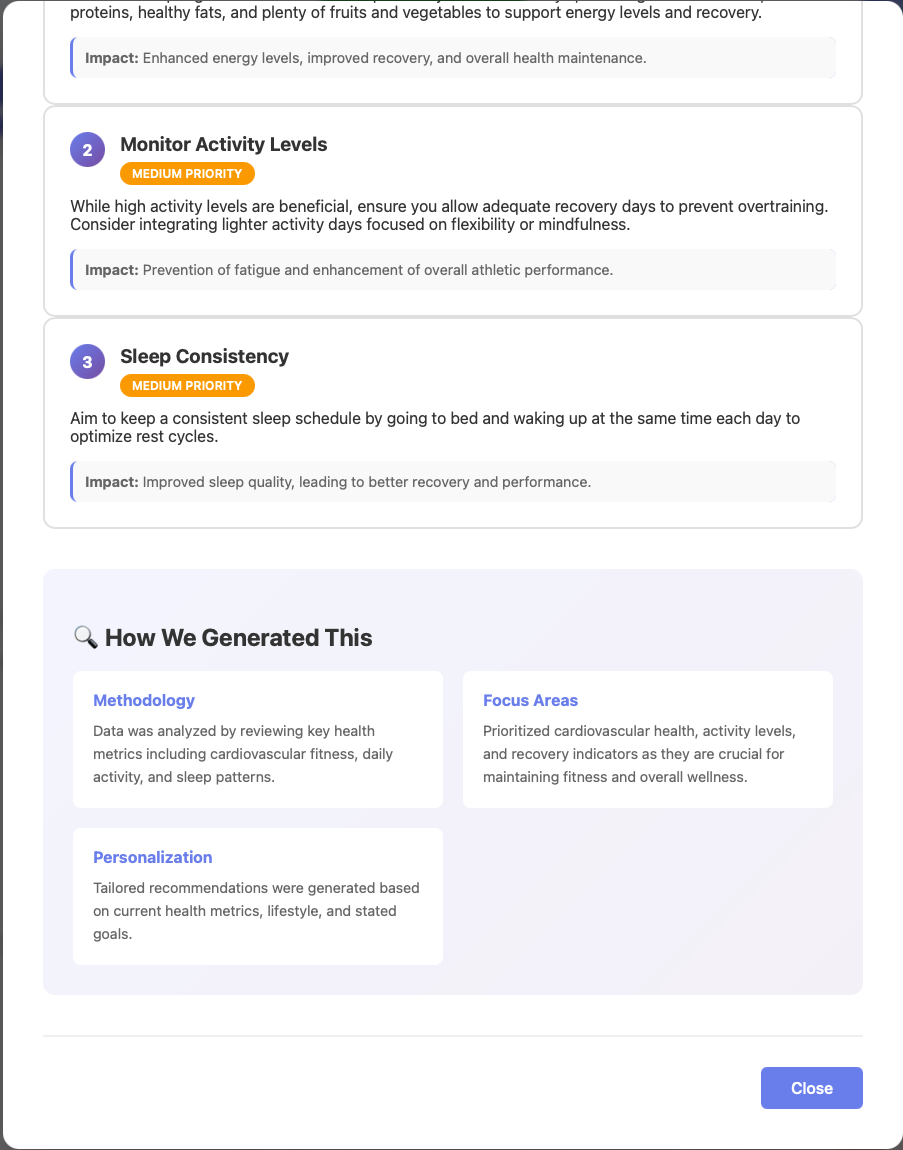}
    \caption{%
      Lower section of Elena's health report showing the prioritised recommendations and explainability panel.
      The high-priority recommendation addresses nutrition, a domain not represented by any sensor metric, illustrating that the report module reasons from lifestyle context rather than solely from measured physiological indicators.
    }
    \label{fig:elena-report-bottom}
          \Description{elena report}

  \end{subfigure}
  \caption{%
    Health report generated for Elena.
    This case is the most analytically informative in the corpus: all physiological metrics are in the excellent or good range, yet the report produces three differentiated recommendations covering nutrition, overtraining risk management, and sleep schedule consistency.
    The explainability panel frames the analysis around maintenance and forward-looking caution rather than corrective action, demonstrating that the report module adjusts its reasoning strategy to match the severity structure of the user's health profile.
  }
  \label{fig:elena-report}
        \Description{Elena report}

\end{figure*}

\paragraph{Elena: recommendations under a healthy baseline.}
Elena's report is the most analytically informative case in the corpus because it demonstrates how the system reasons when all primary physiological indicators are in the excellent range.
Her key metrics section rates VO2 max (68\,ml/kg/min), resting heart rate (44\,bpm), daily steps (18,430), and heart rate variability (112\,ms) as Excellent, and sleep duration (8.45\,hrs) as Good (Figure~\ref{fig:elena-report}).
Faced with no threshold violations, the model does not default to a generic positive summary; instead it generates three substantive recommendations: maintaining balanced nutrition to support energy levels and recovery (high priority), monitoring activity volume to prevent overtraining and incorporating recovery days (medium priority), and establishing sleep schedule consistency to optimise rest cycles (medium priority).
The first recommendation addresses a nutritional gap not directly represented in any sensor metric; the second is prospective risk management rather than a reaction to an observed problem.
The explainability panel identifies cardiovascular health, activity levels, and recovery indicators as focus areas, acknowledging that these are already strong and therefore warrant maintenance and forward-looking caution rather than corrective action.
This case illustrates that the report module can produce clinically meaningful output even when standard threshold-based flags would be silent, by reasoning from goal context and lifestyle patterns rather than from abnormal readings alone.

\paragraph{Goal-driven metric ordering: Tyler.}
Tyler's report provides the clearest demonstration of goal-driven report structure.
Whereas all other users' reports open the key metrics section with VO2 max or steps, Tyler's first card is sleep duration (5.9\,hrs, rated \textit{Needs Attention}), directly reflecting his stated goal of improving sleep.
The remaining cards cover VO2 max (49\,ml/kg/min, \textit{Good}), resting heart rate (62\,bpm, \textit{Excellent}), and heart rate variability (82\,ms, \textit{Excellent}), and the report produces only two recommendations: improving sleep hygiene (high priority) and a physical activity adjustment oriented toward stress reduction to support better sleep (medium priority).
The explainability panel states explicitly that sleep was prioritised in the analysis given its critical role for a young adult.
The consequence of Tyler's strong cardiovascular baseline is that the report can focus almost entirely on his declared problem rather than distributing attention across multiple health dimensions, producing the most tightly scoped output in the corpus.

\paragraph{Multi-flag severity: Priya.}
Priya's report is distinctive in two respects.
First, it is the only report to include a \textit{Concerning} badge, assigned to her self-reported high stress level, which appears as a dedicated metric card alongside the standard physiological indicators.
Second, it is the only report to generate two simultaneous high-priority recommendations: improving sleep quality (6.25\,hrs rated \textit{Needs Attention}) and incorporating stress management techniques.
The co-occurrence of sleep and stress flags in the high-priority tier is consistent with the bidirectional relationship between these two variables, and the model's explainability panel identifies both as primary focus areas, noting their compounding effect on physical health and fitness goal adherence.
The third recommendation, increasing physical activity variety, receives medium priority and is oriented toward long-term fitness diversification rather than acute symptom management.

\paragraph{Age-calibrated assessment: James.}
James's report illustrates a subtler aspect of the model's contextual reasoning: the calibration of status labels to the user's age.
A VO2 max of 29\,ml/kg/min would be rated \textit{Poor} or at best \textit{Below Average} on a population-wide scale for an adult in their mid-forties, yet James's report assigns it a \textit{Good} badge with an interpretive note that this value is typical for his age group and indicative of moderate aerobic fitness.
Similarly, a resting heart rate of 71\,bpm receives an \textit{Excellent} badge, reflecting that this reading is cardiovascularly healthy for a 67-year-old sedentary male, rather than applying a threshold calibrated to younger or more active populations.
The practical consequence is that James's recommendations are framed as realistic and age-appropriate incremental goals (increasing steps to 10,000, incorporating strength training) rather than as corrections to an alarming deficit, which is the appropriate framing for a user whose primary obstacle is habitual sedentariness rather than physiological impairment.
\subsection{Cross-User Analysis}

The per-user narratives in Sections~5.3 and 5.4 document adaptation outcomes individually; this subsection synthesises those observations into cross-user claims about the architecture's behavior.
The evaluation is analytical rather than inferential: with six users and no randomised conditions, it cannot support statistical conclusions about adaptation quality or user satisfaction.
What it can support are claims of a narrower but reproducible kind: whether specific adaptation mechanisms operated as designed, whether their outputs are consistent with the behavioral signals that triggered them, and whether the system's reasoning differentiates appropriately across profiles.
Table~\ref{tab:cross-user} summarises the key adaptation dimensions for each user.

\begin{table*}[t]
  \caption{
    Cross-user adaptation summary across five dimensions.
    \textit{Constraint enforcement} refers to whether explicit widget dislikes were excluded from the generated layout.
    \textit{Priority preservation} refers to whether drag-promoted metrics occupied leading grid positions.
    \textit{Dwell expansion} refers to whether widgets accumulating sustained hover time were widened.
    \textit{Goal alignment} refers to whether the layout and report content reflect the user's stated health objective.}
  \label{tab:cross-user}
  \footnotesize
  \begin{tabular*}{\textwidth}{@{\extracolsep{\fill}} l c p{2.8cm} p{2.8cm} c c @{}}
    \toprule
    \textbf{User} &
    \textbf{Active} &
    \textbf{Constraint} &
    \textbf{Priority} &
    \textbf{Dwell} &
    \textbf{Goal} \\
    & \textbf{Channels} &
    \textbf{Enforcement} &
    \textbf{Preservation} &
    \textbf{Expansion} &
    \textbf{Alignment} \\
    \midrule
    Sarah  & 1 & \checkmark\ (2/2 dislikes)   & N/A                              & N/A         & \checkmark \\[2pt]
    Elena  & 3 & \checkmark\ (1/1 dislike)    & \checkmark\ (drag top 3 matched) & \checkmark\ (3 widgets widened) & \checkmark \\[2pt]
    James  & 2 & N/A                           & \checkmark\ (steps led)          & \checkmark\ (HR content expanded) & \checkmark \\[2pt]
    Marcus & 3 & \checkmark\ (1/1 dislike)    & \checkmark\ (user override) & \checkmark\ (2 widgets widened) & \checkmark \\[2pt]
    Priya  & 2 & N/A                           & \checkmark\ (activity metrics led) & \checkmark\ (stress content expanded) & \checkmark \\[2pt]
    Tyler  & 0 & N/A                           & N/A                              & N/A         & \checkmark \\
    \bottomrule
  \end{tabular*}
\end{table*}

\paragraph{Constraint enforcement is categorical.}
Across the three users who provided explicit widget dislikes, all four dismissed metrics are absent from the generated layout with no exceptions.
Sarah's step count and resting heart rate, Elena's VO2 max, and Marcus's sleep quality breakdown are all excluded regardless of their relevance to the user's goal profile.
This categorical behaviour is by design: explicit feedback is encoded as a hard constraint in the prompt rather than a weighted preference, and the evaluation confirms that this constraint survives the synthesis process even when the disliked metric would otherwise be goal-appropriate.
The most informative instance is Elena's dislike of VO2 max, a metric that her athletic profile would ordinarily make central; its absence from the layout with no substitution demonstrates that the hard-constraint mechanism does not negotiate with goal-level reasoning.

\paragraph{Drag and dwell operate on orthogonal layout dimensions.}
A consistent pattern across all four users who generated both spatial and dwell signals is that the two channels influence different properties of the layout without interfering with each other.
Drag reorganisation affects the ordering of widgets: the metrics that a user manually positions at the top of the grid appear at the top of the regenerated layout, preserving their hierarchical prominence.
Dwell time affects the sizing of widgets: metrics that accumulate sustained hover duration are rendered at wide size rather than normal size, expanding their visual real estate without necessarily elevating their position.
In Elena's layout, exercise minutes, HRV, and hourly stress occupy the top positions through drag priority, while the weekly steps chart, recent activities timeline, and anxiety heatmap are wide through dwell expansion; no widget in her layout is both top-positioned and widened, illustrating the orthogonal contribution of each channel.
This separation is architecturally coherent: it means the two signals can be simultaneously active without producing conflicting instructions, and it allows the rendering subsystem to honour both without trade-offs.

\paragraph{Profile-driven adaptation is sufficient but less specific.}
Comparing Tyler, who received no behavioral signals, with the five users who did, reveals the marginal contribution of the behavioral channels.
Tyler's layout is structurally coherent with his stated goal: sleep metrics appear prominently, a sleep trend chart occupies a wide position, and no contradictory content is present.
However, the layout contains no features that reflect individual preference or usage patterns: every widget size is at the default setting, every position is AI-assigned, and no metric is excluded beyond what the temporal context and goal profile would ordinarily suppress.
By contrast, the layouts of users who generated signals contain explicit traces of individual behavior: absent metrics, user-selected leading positions, and expanded widgets whose widening can be attributed to a specific dwell event rather than to generic design logic.
The gap between Tyler's layout and Elena's or Marcus's is not a quality gap in any absolute sense; it is a specificity gap, and that gap is entirely attributable to the behavioral tracking channels.

\paragraph{Report reasoning differentiates across the full profile spectrum.}
The cross-user report analysis in Section~5.5 provides a parallel set of observations about the model's goal-sensitive reasoning in a different output modality.
Across the five users for whom report content is interpretable, no two reports produce the same recommendation structure: the number of recommendations ranges from two (Tyler) to three (all others), the distribution of priority badges differs, and the metric cards selected for analysis vary both in identity and in status label.
The most diagnostic finding is that status labels are not applied uniformly: a VO2 max of 29\,ml/kg/min is rated \textit{Good} for James but would be rated far lower for a younger, more active user, indicating that the model applies contextually calibrated thresholds rather than fixed population norms.
Similarly, the system generates a \textit{Concerning} badge for Priya's stress level and two simultaneous high-priority flags, while generating only maintenance-oriented recommendations for Elena despite her higher overall activity load.
Taken together, the dashboard and report outcomes suggest that the same language model, operating under different structured prompt inputs, is capable of producing qualitatively differentiated outputs that are coherent with the individual user's goal, metric context, and interaction history across the full range of profiles in this evaluation.

\paragraph{Scope and limitations.}

The corpus covers six users, which is sufficient for analytical demonstration but not for claims about generalisability across user populations.
All sessions were conducted in a single temporal phase per user, so time-of-day adaptation, while encoded in the prompt architecture, is not directly evaluated across the sessions reported here.
The behavioral signal interpretations applied by the system, that prolonged hover indicates interest warranting detail and that top-positioned drag targets indicate priority, are heuristic approximations whose validity for individual users cannot be confirmed without additional feedback or longitudinal follow-up.
Finally, the evaluation assesses whether the adaptation mechanisms operated as designed, not whether the resulting interfaces improved user experience, reduced cognitive load, or supported better health outcomes; establishing those effects would require a controlled user study with outcome measures extending beyond a single session.
\color{black}
\section{Discussion}

\subsection{Technical Contributions}

This paper contributes three engineering results embodied in a working prototype rather than completed empirical claims of adaptive benefit.
First, the multi-modal behavioral aggregation architecture demonstrates how heterogeneous user signals can be tracked asynchronously through independent monitoring channels and synthesized during adaptation without requiring a unified behavioral model.
Traditional adaptive systems typically define a single behavioral representation (e.g., user model or preference vector) that all signals must update, creating tight coupling between signal types and adaptation logic.

Our approach maintains loose coupling by treating each signal type independently during collection and performing synthesis only during prompt construction, enabling extensibility: new signal types can be added by implementing additional tracking modules and extending prompt templates without modifying core adaptation logic.

Second, the structured prompt engineering methodology provides a generalizable approach for using LLMs as behavioral synthesis engines rather than endpoint generators. The hierarchical prompt structure with separated sections for temporal context, behavioral signals, explicit preferences, and user profiles enables the LLM to reason about multiple constraint types simultaneously while maintaining generation consistency across adaptation cycles. The semantic bridging approach-explicitly stating the interpretation of behavioral patterns rather than expecting the LLM to infer meaning from raw data-proves critical for reliable adaptation. This methodology can be applied beyond health dashboards to any adaptive interface domain where designers can articulate how different behavioral signals should influence interface decisions.
This mechanism positions the LLM as a synthesis component operating over explicitly defined inputs and heuristics rather than as an unconstrained interface generator.

Third, the integrated explainability generation mechanism demonstrates how LLM-based adaptive systems can provide transparency without separate explanation modules. By requesting explanatory text as part of the layout generation output using the structured schema with suggestion, nudge, and meta-explanation fields, the system obtains human-readable rationale for adaptations as a byproduct of the generation process itself. This approach avoids the complexity of post-hoc explanation generation techniques that must reconstruct reasoning from final outputs, instead capturing adaptation logic directly from the reasoning process. The three-tier explanation model (actionable suggestion, scientific rationale, meta-explanation of design choices) provides transparency~\cite{Doshi2017,Miller2019} at multiple levels appropriate for different user information needs.

These elements together define a reproducible architectural pattern for LLM-mediated adaptive interfaces in settings where sparse behavioral signals, evolving context, and interpretable regeneration are more realistic than large-scale historical training data.

\subsection{Architectural Limitations and Engineering Challenges}

The current architecture exhibits several technical limitations that suggest directions for future engineering research. The system relies on OpenAI's commercial API for LLM access, introducing observed latency (typically five to eight seconds for layout generation), cost (approximately point two cents per regeneration), and dependency on external service availability. These factors may prove acceptable for applications requiring only occasional adaptation but would be problematic for interfaces requiring immediate response to every user action. Future work might explore distillation approaches where a large commercial model generates training data for a smaller locally-hosted model that can perform inference with lower latency and cost, or investigate caching strategies that reuse layout components across similar contexts.

The prompt engineering approach, while effective in practice, lacks formal guarantees about output validity. Despite explicit instructions and schema specifications, LLMs can occasionally generate layouts violating technical constraints, such as specifying non-existent data sources or incompatible widget type and data source combinations. The current implementation handles these failures through validation and fallback to previous valid layouts, but more robust approaches might employ constrained generation techniques that syntactically prevent invalid outputs, or implement iterative refinement where the system detects violations and requests corrections from the LLM. Research on formal verification of LLM-generated interface specifications could substantially improve reliability for production deployment.

The behavioral interpretation strategies employ relatively simple heuristics that may not generalize across all users or contexts. For example, the attention allocation module interprets sustained hover as analytical interest, but some users might exhibit prolonged hover due to confusion, distraction, or accessibility needs requiring longer processing time. More sophisticated interpretation could incorporate additional signals such as cursor movement patterns within widgets, scrolling behavior suggesting continued reading versus abandonment, or temporal patterns distinguishing focused study from repeated confusion-driven revisiting. Machine learning approaches could discover more nuanced behavioral interpretations from user satisfaction signals, though this would reintroduce the cold start problem that the LLM-based approach avoids.

A further limitation concerns the absence of an internal mechanism for evaluating whether the adaptations the system produces actually improve the user experience. The system reads behavioral signals such as dwell time and drag reorganization, interprets them as indicators of interest and priority, and acts on them in the next regeneration cycle. What the architecture does not currently include is a way of checking whether those adaptations were perceived as useful by the user. If a behavioral signal is interpreted incorrectly, for example a prolonged hover that reflects confusion rather than interest, the system will continue adapting in that direction until the user issues an explicit like or dislike. This means that, in the present prototype, explicit feedback is the only corrective signal available, and subtle misinterpretations of implicit signals may persist across sessions without being surfaced.
Addressing this limitation properly requires moving beyond what an artifact-centered evaluation can show. Whether the adaptations make the dashboard genuinely easier to use, more informative, or more engaging is a question about user experience, and it cannot be answered by inspecting layouts and explainability objects alone. We therefore plan to follow this work with a longitudinal user study in which participants interact with the system over multiple sessions, with empirical measures of usability, perceived helpfulness, and adaptation accuracy collected throughout. A study of that kind is, in our view, the only way to establish whether the architecture's behavioral interpretations are aligned with how users actually engage with their dashboards over time, and it is the natural next step once the engineering contribution presented here has been laid out in full.

\subsection{Generalization Beyond Health Dashboards}

While our implementation targets health dashboards, the architectural approach generalizes to any adaptive interface domain meeting three conditions. First, the domain must involve presenting information from multiple sources where users exhibit heterogeneous preferences about importance and presentation modality. This condition ensures adaptation has meaningful design space to explore rather than just switching between a few predefined configurations. Second, the domain must enable articulation of semantic interpretations for behavioral patterns that an LLM can reason about. This requires that designers can explain in natural language how specific user actions should influence interface decisions, enabling construction of the bridging statements in prompts. Third, the domain must allow sufficient adaptation latency to accommodate LLM generation time, or adaptation must occur infrequently enough that occasional delays are acceptable.

Promising application domains beyond health dashboards include: development environment customization where IDEs adapt tool layouts, panel visibility, and shortcut configurations based on observed workflow patterns and explicit preferences; business intelligence dashboards where metric prominence and visualization types adjust based on role, current tasks, and analytical focus patterns; smart home control interfaces where device group organization and automation suggestions adapt to usage patterns and temporal contexts; and educational platform interfaces where content recommendations, difficulty progression, and support resource visibility adapt to learning patterns and expressed preferences. Each domain would require domain-specific prompt engineering to encode appropriate contextual knowledge and behavioral interpretations, but the core architectural pattern of asynchronous multi-modal tracking synthesized through structured LLM prompting would transfer directly.

\subsection{Future Technical Directions}

Several technical extensions would enhance the architecture's capabilities and applicability. Incorporating explicit feedback loops where users can correct or refine adaptations would enable continuous prompt engineering refinement. The system could ask users whether specific adaptations were helpful and use these signals to adjust behavioral interpretation strategies or prompt weighting parameters over time. This meta-learning approach could discover domain-specific best practices for behavioral synthesis automatically rather than requiring manual prompt engineering.

Exploring collaborative filtering approaches where adaptations leverage aggregated behavioral patterns across multiple users with similar profiles could accelerate personalization, particularly during cold start periods before substantial individual behavioral history accumulates. The LLM could receive prompts incorporating insights like "users with similar weight loss goals and stress levels typically find these metrics most valuable during evening sessions" to inform initial adaptations. Privacy-preserving aggregation techniques would be essential to enable cross-user learning while protecting individual data.

Investigating reinforcement learning integration where the system treats layout generation as a policy and learns from implicit satisfaction signals (session duration, regeneration frequency, metric interaction patterns) could enable continuous improvement of adaptation quality. The LLM-generated layouts would serve as initial policies that the learning system refines over time, combining LLM common sense reasoning with data-driven optimization. This hybrid approach could potentially achieve better adaptation quality than either technique alone.

Finally, extending the architecture to handle multi-device adaptation where layouts adjust not only to user preferences and temporal context but also to device form factors and usage contexts (smartphone during commute, tablet during leisure, desktop during focused work) would increase practical utility. The structured prompt engineering approach naturally accommodates additional contextual factors, suggesting straightforward extension to device-aware adaptation.

\section{Conclusion}

We have presented a technical architecture for continuous behavioral synthesis in adaptive user interfaces using Large Language Models as reasoning engines. The architecture addresses the fundamental challenge of translating heterogeneous low-level user interactions into coherent high-level interface design decisions through structured prompt engineering that makes behavioral semantics explicit. Our approach enables effective adaptation without requiring extensive training data by leveraging LLM embedded knowledge about interface design principles and domain-specific conventions, while maintaining transparency through integrated explainability generation.

The multi-modal behavioral tracking subsystem demonstrates how explicit feedback, spatial reorganization patterns, and attention allocation signals can be monitored asynchronously and synthesized during adaptation without requiring a unified behavioral model. The structured prompt engineering methodology provides a generalizable approach for behavioral synthesis that separates temporal context, behavioral interpretation, explicit preferences, and user profiles into distinct prompt sections that the LLM can reason about simultaneously. The widget-based rendering architecture enables clean separation between adaptive logic and interface implementation, supporting extensibility and modification without impacting core adaptation mechanisms.

Validation through a complete implementation in the health dashboard domain demonstrates that the approach produces coherent adaptive behavior that respects user preferences while exercising design judgment about presentation details. The system generates explanatory text revealing adaptation rationale, supporting user understanding~\cite{Liao2020,Abdul2018} and trust~\cite{Kulesza2013,Amershi2019} in automated personalization. The technical approach generalizes beyond health dashboards to any adaptive interface domain where behavioral patterns can be articulated semantically and where adaptation latency accommodates LLM generation time.

This work contributes to the broader research agenda of engineering intelligent interactive systems by demonstrating how Large Language Models can serve as behavioral synthesis components within adaptive interface architectures. As LLMs become increasingly integrated into interactive systems, architectural patterns that combine their reasoning capabilities with principled behavioral learning will be essential for creating truly personalized user experiences that adapt continuously while maintaining transparency and respecting user agency.

\bibliographystyle{ACM-Reference-Format}
\bibliography{references}

\end{document}